\documentclass[twocolumn,final,epj]{svjour}
\pdfoutput=1
\usepackage[utf8]{inputenc}
\usepackage[T1]{fontenc}
\usepackage{graphicx}
\usepackage{amssymb}
\usepackage{amsmath}
\usepackage{hyperref}

\def\pdag{{\phantom\dagger}}
\newcommand{\bra}[1]{\left\langle #1 \right|}
\newcommand{\ket}[1]{\left| #1 \right\rangle}

\begin{document}

\title{Dynamical decoherence of a qubit\\
coupled to a quantum dot or the SYK black hole}

\author{
Klaus M. Frahm\inst{1} \and 
Dima L. Shepelyansky\inst{1}}

\institute{
Laboratoire de Physique Th\'eorique, IRSAMC, 
Universit\'e de Toulouse, CNRS, UPS, 31062 Toulouse, France
}

\titlerunning{Dynamical decoherence of qubit}
\authorrunning{K.M.~Frahm and D.L.~Shepelyansky}

\date{Dated: 28 April 2018}

\abstract{We study the dynamical decoherence of a qubit weakly
coupled to a two-body random interaction model (TBRIM)
describing a quantum dot of interacting fermions
or the Sachdev-Ye-Kitaev (SYK) black hole model.  
We determine the rates of qubit relaxation
and dephasing for regimes of dynamical thermalization
of the quantum dot or of quantum chaos in the SYK model.
These rates are found to correspond to the Fermi golden rule 
and quantum Zeno regimes depending on the qubit-fermion coupling strength.
An unusual regime is found where these rates are practically independent of
TBRIM parameters. We push forward an analogy between TBRIM
and quantum small-world networks with an explosive
spreading over exponentially large number of states in a finite time
being similar to six degrees of separation in small-world social networks.
We find that the SYK model has approximately two-three degrees of separation.
}


\maketitle

\section{Introduction}

The problem of qubit decoherence is crucial for 
the process of quantum measurement \cite{braginsky} and
the field of quantum information and computation \cite{chang}.
The experimental realization of superconducting qubits
\cite{nakamura,esteve} extended this problem to
a world of large objects due to
a macroscopic size of superconducting qubits
(see e.g. \cite{averin,shnirman,wendin}).
In theoretical considerations the decoherence of a qubit is 
usually due to the contact with a thermal bath,
weak measurements or other statistical systems 
characterizing a detector (or sensor) being in a contact
with the qubit \cite{averin,shnirman,wendin}.
A model of a deterministic detector, whose evolution 
takes place in a regime of quantum chaos, 
was studied in \cite{lee2005}
demonstrating the emergence of dynamical decoherence of a 
qubit in absence of any thermal bath, noise and external randomness.
We extend this research line \cite{lee2005}
considering as a deterministic detector
a quantum dot with interacting fermions or
the Sachdev-Ye-Kitaev (SYK) black hole model.

The question about dynamical decoherence is closely related 
to the problem of quantum dynamical thermalization 
and random matrix theory (RMT) invented by Wigner \cite{wigner,mehta,guhr}
for the description of complex atoms and nuclei. 
While the properties
of one-particle quantum chaos and their link with RMT
are now mainly understood (see e.g. \cite{bohigas1984,haake,ullmo}),
the analysis of many-body quantum systems is
more difficult due to the complexity
of quantum many-body systems (QMBS). Furthermore RMT 
is only an approximation to QMBS since in nature 
we have only two-body interactions and hence the
exponentially large Hamiltonian matrix of QMBS 
has only a small fraction of non-zero matrix elements.
To capture this feature a two-body random interaction model 
of fermions (TBRIM)
was proposed in \cite{french1,bohigas1,french2,bohigas2}
and it was shown that at strong interactions 
this model is characterized by RMT level spacing statistics.
The first numerical results and analytical arguments
for a critical interactions strength in TBRIM with a
finite level spacing $\Delta$ between one-particle orbitals
was proposed by Sven {\AA}berg in \cite{aberg1,aberg2}.
For the TBRIM the {\AA}berg criterion for onset of quantum
chaos and dynamical thermalization has the form
\begin{equation}
\label{eq:abergcriterion}
\delta E = E - E_g > \delta E_{\rm ch} \approx g^{2/3} \Delta \;, \;\; g = \Delta/U  ,
\end{equation}
where $U$ is a typical strength of two-body interactions,
$\Delta$ is an average one-particle level spacing in 
a finite size quantum dot with interacting fermions,
$E_g$ is the ground state energy of the quantum dot
when all electrons are below the Fermi energy $E_F$ and 
$E$ is the energy of an excited eigenstate. The dimensional
parameter $g \gg 1$ is assumed to be large
playing the role of the conductance of a quantum dot
with weakly interacting electrons. 
The validity of the {\AA}berg criterion (\ref{eq:abergcriterion}) for the 
emergence of RMT level statistics was confirmed in first numerical
simulations \cite{aberg1,aberg2} 
and in independent more extensive analytical and
numerical studies for 
3 particles in a quantum dot \cite{sushkov},
TBRIM \cite{jacquod}, spin glass shards \cite{georgeotspin},
quantum computers with imperfections \cite{georgeotqc,nobel,benentiqc}.
Advanced theoretical arguments developed
in \cite{mirlin1,mirlin2} confirm
the relation (\ref{eq:abergcriterion}) 
for interacting fermions in a quantum dot.

While the validity of the {\AA}berg criterion
for emergence of RMT in TBRIM and other models
is satisfactory confirmed by numerical and analytical studies,
a dynamical thermalization conjecture (DTC),
which is used for the derivation of (\ref{eq:abergcriterion}),
is more difficult for the numerical verification
since it requires the knowledge not only of
the eigenvalues but also the computation of eigenstates
that is more difficult. 
The TBRIM numerical results \cite{flambaum}
for the probability distribution
over one-particle orbitals,
averaged over many random realizations,
showed a certain proximity to the Fermi-Dirac distribution
expected from the quantum statistical mechanics \cite{landau}. 
The validity of the Fermi-Dirac distribution
for a single eigenstate was demonstrated
numerically for eigenstates of a quantum computer with 
imperfections and residual inter-qubit couplings \cite{benentiqc}.
We stress that the DTC is proposed for a purely isolated 
system without any contact to an external thermostat and the dynamical
thermalization is only due to internal many-body quantum chaos.
 
However, for a single eigenstate the fluctuations of 
probabilities $n_k$ on one-particle orbitals
are significant requiring heavy large matrix diagonalizations
to obtain a reasonable agreement with the Fermi-Dirac distribution 
\cite{benentiqc}.
Another method was developed for nonlinear disordered
chains described by classical Hamiltonian equations
\cite{mulansky,ermannnjp}. It is based on the computation
of entropy $S$ and energy $E$ tracing the dependence $S(E)$
which is obtained as an implicit function from $S(T)$ and $E(T)$
where $T$ is the system temperature appearing due to dynamical thermalization
in a completely isolated system without any contact to an external thermostat. 
Since the quantities $S$ and $E$ are extensive \cite{landau}
their fluctuations are reduced due to self-averaging. 
The dependence $S(E)$ for many-body quantum systems 
was computed for bosons in disordered Bose-Hubbard model in 1D 
\cite{schlageck} and for spinless fermions in the TBRIM \cite{kolovsky2017}.
These studies demonstrated the stability and efficiency
of $S(E)$-computations confirming validity of the DTC for many-body  
interacting quantum systems. 
The dynamical thermalization of an individual eigenstate 
was also demonstrated in \cite{schlageck,kolovsky2017}.
At present the interest of 
many-body interacting quantum systems is also growing
in the context of many-body localization (MBL) 
and the eigenstate thermalization hypothesis (ETH)
(see e.g. \cite{huse,polkovnikov,borgonovi,alet}).

Another bust of interest to the TBRIM type models
appeared due to the recent results 
of Sachdev-Ye-Kitaev
for a strange metal and its links to 
a quantum black hole model in 1+1 dimensions (coordinate plus time)
known now as the SYK black hole
\cite{sachdevprl,kitaev,sachdevprx}. In fact, the SYK model,
in its fermionic formulation,
corresponds to the TBRIM considered 
in the limit of very strong interactions with a conductance close to zero
$g \rightarrow 0$. The analogy between 
physical representations of the SYK model
attracted a significant interest
of researchers in quantum gravity, many-body systems, RMT
and quantum chaos (see e.g. \cite{rosenhaus,maldacena,tezuka}).
Recent advanced numerical and analytical results on the validity of 
RMT for the SYK model with Majorana fermions
are reported in \cite{garcia1,garcia2,wettig}.

In this work we study the dynamical decoherence 
of a qubit coupled to the TBRIM model.
This is a completely isolated system
in absence of noise, thermal bath and 
external decoherence. At $g \gg 1$ 
the qubit is coupled to a quantum dot
of weakly interacting fermions 
with our main interest being focused on the regime
of dynamical thermalization when the {\AA}berg
criterion (\ref{eq:abergcriterion})
is satisfied. At  $g \ll 1$ our model becomes
equivalent to the SYK black hole model with a qubit 
coupled to it. We note that the decoherence of a 
qubit coupled to a quantum black hole
is extensively discussed in the context of 
the black hole problem of information loss for the 
infalling observer (see \cite{harlow} and Refs. therein).
We expect that the dynamical qubit decoherence 
considered here will be useful for a better
understanding of this problem.

The paper is composed as follows: 
In Section 2 the TBRIM is introduced and some of its properties are reminded 
while in Section 3 the additional qubit-fermion coupling is introduced. 
The qubit relaxation rates are studied in Section 4 and in Section 5 
the link to a quantum small-world networks is discussed. In Section 6
results of the residual level of qubit density matrix relaxation at long 
times are described and Section 7 concludes with the discussion. 
In Appendix A a rather detailed analytical and numerical study for 
the approximate Gaussian form of the average density of states of the TBRIM 
is presented while Appendices B and C deal with the specific issues 
of weakly excited initial states of the TBRIM, where it is difficult to 
obtain clear relaxation rates, and initial states with negative temperatures. 

\section{TBRIM construction and properties}

As in Ref. \cite{kolovsky2017} 
we consider the TBRIM \cite{jacquod}  
with $M$ one-particle orbitals and $0\le L\le M$ spinless fermions
with the Hamiltonian:
\begin{equation}
\label{eq_TBRIM}
H_I=\frac{1}{\sqrt{M}}\sum_{k=1}^M v_k\,c^\dagger_k c^\pdag_k
+\frac{4}{\sqrt{2M^3}}\sum_{i<j,k<l} J_{ij,kl}\,c^\dagger_i c^\dagger_j 
c^\pdag_l c^\pdag_k
\end{equation}
where $c^\dagger_k$, $c^\pdag_k$ are fermion operators for the $M$ orbitals 
satisfying the usual anticommutation relations. Here $v_k$ ($J_{ij,kl}$) 
are real Gaussian random variables with zero mean and variance 
$\langle v_k^2\rangle=V^2$ 
($\langle J_{ij,kl}^2\rangle=J^2(1+\delta_{ik}\delta_{jl})$) such that 
the non-interacting orbital one-particle energies are given by 
$\epsilon_k=v_k/\sqrt{M}$. The variance of the interaction matrix 
elements is chosen such they correspond to a GOE-matrix 
(Gaussian orthogonal ensemble) of 
size $M_2\times M_2$ with $M_2=M(M-1)/2$. The number of 
nonzero elements for a column (or row) of $H_I$ is 
$K = 1 + L(M-L) + L(L-1)(M-L)(M-L-1)/4$ \cite{jacquod,flambaum}.

As shown in Appendix \ref{appa} the density of 
states (DOS) of the TBRIM Hamiltonian (\ref{eq_TBRIM}) 
is approximately Gaussian 
\begin{equation}
\label{eq_DOS_TBRIM}
\rho(E)\approx\frac{d}{\sqrt{2\pi\sigma^2}}\,
\exp\left(-\frac{E^2}{2\sigma^2}\right)
\quad,\quad 
\sigma=\sqrt{\frac{L(M-L)}{M(M-1)}}\,V_{\rm eff}
\end{equation}
which is normalized to 
$d=M!/(L!(M-L)!)$ being the dimensionality of the Hilbert space 
for $M$ orbitals and $L$ particles and 
\begin{equation}
\label{eq_Veff}
V_{\rm eff}=\sqrt{V^2+a(M,L)\,J^2}
\end{equation}
is a rescaled effective energy scale taking into account the increase of 
$\sigma$ due to finite values of $J$. 
The coefficient $a(M,L)$ is computed in Appendix \ref{appa} 
from the average of $\langle\mbox{Tr}(H_I^2)\rangle$ with the result~: 
\begin{equation}
\label{eq_a_coeff}
a(M,L)=\frac{2(M-1)(L-1)}{M^2}
\left(\frac{4}{M-L}+M-L+3\right). 
\end{equation}
The expression (\ref{eq_DOS_TBRIM}) 
fits numerically quite well the DOS for sufficiently large values 
of $M$ and $L$ and even in the 
SYK-case, i.e. when $J\neq 0$ but $V=0$, it is quite accurate. 
The corresponding average many body level spacing (at the band center) is 
$\Delta_{\rm MB}=\sqrt{2\pi}\,\sigma/d$. For later use we also define 
an effective rescaled average one-particle level spacing 
by $\Delta_1=\sqrt{2\pi}V_{\rm eff}/M^{3/2}$. At $J\ll V$ we have 
$V_{\rm eff}\approx V$ and $\Delta_1$ is just the average distance of 
the one-particle energies $\epsilon_k$ (in the band center). 
Thus the effective dimensionless conductance of 
our TBRIM (see \cite{jacquod})
is $g \approx \Delta_1/U_{s} \approx \sqrt{\pi} V_{\rm eff}/2J 
\approx V/J \gg 1$ for $J \ll V$ and
$g \approx 1$ for $J \gg V$ at $M \approx L/2$
($U_s = 2\sqrt{2} J /M^{3/2}$ is an effective interaction strength).

Since we are using only a small number of 
statistical realizations, we have chosen realizations of $v_k$ such that 
exactly $\sum_k v_k=0$ and $\sum_k \epsilon_k^2=(1/M)\sum_k v_k^2=V^2$. 

We have numerically diagonalized $H_I$ and done further numerical 
computations described below for the cases $M=12$, $M=14$ 
and $M=16$ with $L=M/2-1\approx M/2$. In this work we only show the 
results for the case of largest matrix size $M=16$ and $L=7$ corresponding to 
$d=11440$ (for this case the coefficient in
(\ref{eq_Veff}) and (\ref{eq_a_coeff}) is just $a(16,7)=8.75$
and the number of nonzero matrix elements per row/column of $H_I$ is $K=820$).
Unless stated otherwise, all results presented below, especially 
in the figures apply to this case. We have, however, verified that the 
physical interpretation of the results also apply to the cases of 
smaller matrix size (with some restrictions concerning reduced times 
scales for the long time behavior, more limited parameter range etc.).
We present the results for one specific disorder realization
but we checked that, apart from fluctuations, 
the results remain stable for other realizations.

First we diagonalize numerically one realization of $H_I$ 
for $M=16$, $L=7$, $V=\sqrt{14}\approx3.74166$, various values of $J$ 
or the SYK-case (i.e. $V=0$, $J=1$). Similar to 
\cite{kolovsky2017} we determine for each many body eigenstate the 
occupation numbers 
$n_k=\langle c^\dagger_k c^\pdag_k\rangle$
with the corresponding fermion entropy \cite{landau}~:
\begin{equation}
\label{eq_entropy}
S=-\sum_{k=1}^M\Bigl(n_k\,\ln n_k+(1-n_k)\,\ln(1-n_k)\Bigr)
\end{equation}
and the effective one-particle total energy 
\begin{equation}
\label{eq_1p_energy}
E_{1p}=\sum_{k=1}^M\,\epsilon_k\,n_k
\end{equation}
based on the assumption on non- or weakly-interacting fermions. These energies 
are rather close to the exact many body energies 
$E_{\rm ex}\approx E_{1p}$ provided $J\ll V$. 

\begin{figure}[t]
\begin{center}
\includegraphics[width=0.48\textwidth]{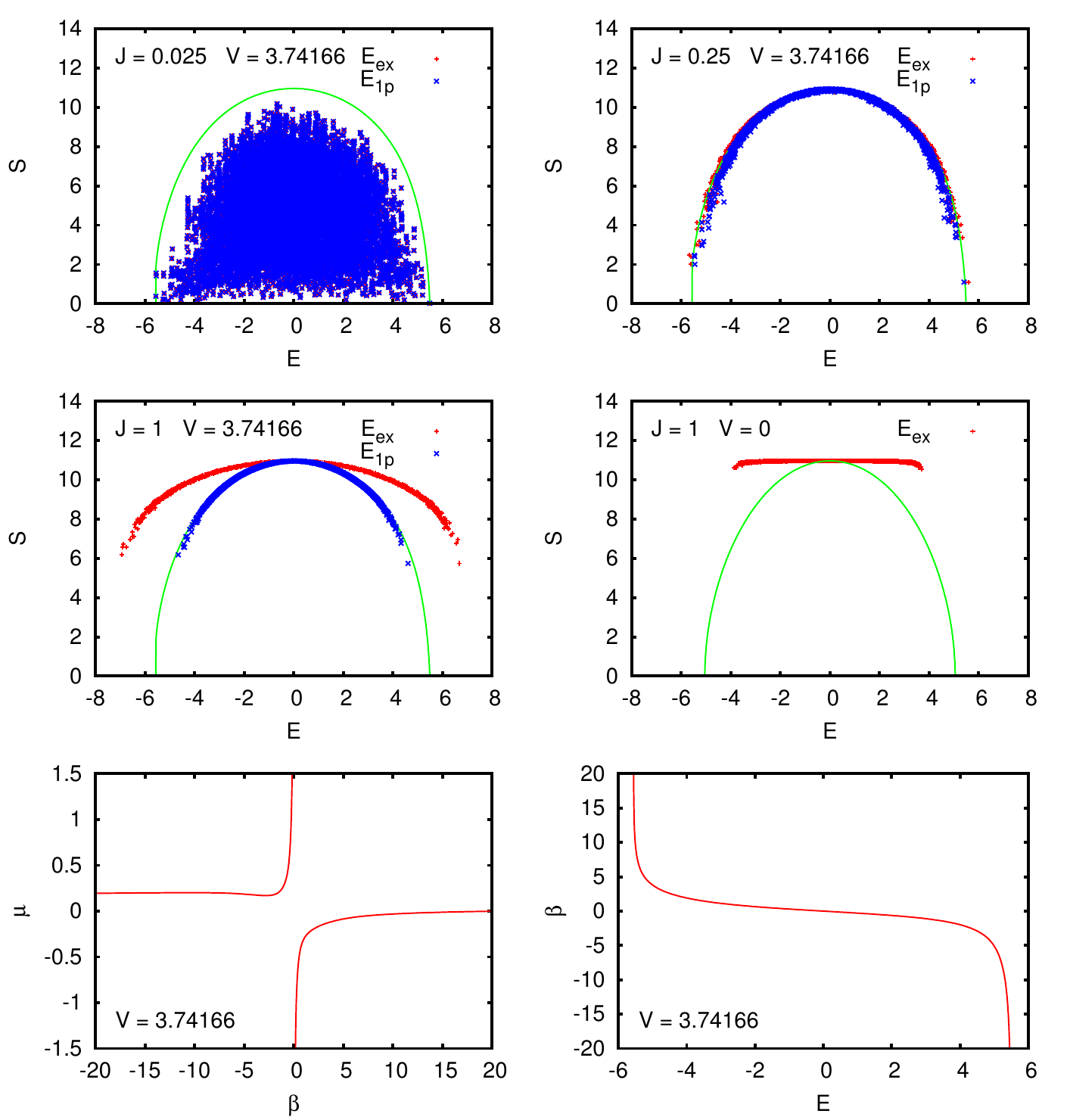}
\end{center}
\caption{\label{fig1}(color online) {\em Top and center panels:} 
Dependence of the fermion entropy $S$ 
given by (\ref{eq_entropy}) on the effective one-particle total energy 
$E_{1p}$ defined in (\ref{eq_1p_energy}) (blue cross symbols) 
and the exact many body energy $E_{\rm ex}$ (red plus symbols). 
The green curve shows the theoretical Fermi-Dirac thermalization ansatz
(\ref{eq_fermidirac}) as explained in the text. 
All panels correspond to $M=16$ orbitals, $L=7$ particles and 
Hamiltonian matrix size $d=11440$. 
Both top and center left panels  correspond to $V=\sqrt{14}\approx 3.74166$ 
and $J=0.025$ (top left), $J=0.25$ (top right) and $J=1$ (center left). 
Center right panel corresponds to the SYK case at $V=0$ and $J=1$ with 
the green curve computed from a model of equidistant one-particle energies 
of non-interacting fermions.
{\em Bottom panels:} 
Dependence of the inverse temperature
$\beta=1/T$ on energy $E$ (bottom right panel) and chemical potential $\mu$
on $\beta$ (bottom left panel) corresponding to the 
Fermi-Dirac ansatz for the set of one-particle energies  $\epsilon_k$
used for the chosen realization of 
$H_I$ at $V=\sqrt{14}\approx 3.74166$. 
}
\end{figure}

In Fig.~\ref{fig1} we 
compare the dependence of $S$ on both energy scales with the theoretical 
fermionic behavior where $n_k$ in (\ref{eq_entropy}) is replaced by 
the usual thermal
Fermi-Dirac distribution (or ansatz) over one-particle orbitals \cite{landau}:
\begin{equation}
\label{eq_fermidirac}
n_k=1/(1+\exp[\beta(\epsilon_k-\mu)]) \;, \;\; \beta=1/T
\end{equation}
with the inverse temperature $\beta$ and chemical potential $\mu$ 
determined by the implicit conditions (\ref{eq_1p_energy}) and 
$L=\sum_k n_k$ with the given set of diagonal one-particle energies 
$\epsilon_k$. For the SYK case with $V=0$ and $J=1$ we choose for the 
``theoretical'' curve the case of one-particles energies equidistant values 
$\epsilon_k$ such that $\sum_k \epsilon_k=0$ and 
$\sum_k \epsilon_k^2=V_{\rm eff}^2$ with the effective rescaled 
energy scale (\ref{eq_Veff}) at $V=0$ and $J=1$ (
a similar procedure was used in \cite{kolovsky2017}
for this SYK case).

At $V=\sqrt{14}\approx 3.74166$ one can observe in Fig.~\ref{fig1} 
the onset of thermalization with increasing interaction strength $J$. 
At very weak interaction $J=0.025$ the entropy is typically below the 
theoretical behavior indicating that the system is not thermalized. We can 
also mention that for this case the level spacing distribution of $H_I$ 
does not obey the Wigner surmise (for the GOE case) and is closer to the 
Poisson distribution (with some small level-repulsion for very short energy 
differences). 
At $J=0.25$ (this value  corresponds to the case $J=1$ 
in \cite{kolovsky2017} due to a difference in the normalization) 
the system is well thermalized but the 
interaction is still sufficiently low so that $E_{1p}\approx E_{\rm ex}$. 
Here and also for larger values of $J$ the level spacing distribution 
clearly corresponds to the Wigner surmise
(this was also seen in \cite{kolovsky2017} and 
we do not show these data here). Thus at $J=0.25$ 
we have onset of the dynamical thermalization induced by weak many-body
interactions.
At $J=1$ the data points for $E_{1p}$ coincide very well with the theoretical 
fermionic curve confirming the onset of dynamical 
thermalization induced by interactions. However, here due the 
stronger interaction values the ratio $E_{\rm ex}/E_{1p}$ is considerably 
larger than unity. 

For the SYK case $V=0$, $J=1$ the entropy is close to 
its maximal value $S\approx 11$ for nearly all eigenstates and the theoretical 
model of equidistant one-particle energies is not confirmed. This value of $S$ 
is actually consistent with $n_k\approx 0.5$ for all orbitals $k$ which gives 
due (\ref{eq_entropy}) $S\approx 16\ln(2)\approx 11.1$. For the SYK case 
the numerical level spacing distribution also corresponds to the Wigner 
surmise. 

The results of this Section show that at moderate
interactions with $g \ll 1$ the DTC is well working (e.g. 
$J=0.25, V=\sqrt{14}, g \approx 15$)
and the dependence $S(E)$ is well described by the thermal Fermi-Dirac
distribution (\ref{eq_fermidirac}). Of course, at very small interactions
(e.g. $J=0.025,\ V=\sqrt{14},\ g \approx 150$) 
the DTC is not valid in qualitative
agreement with the {\AA}berg criterion (\ref{eq:abergcriterion}).
Here we do not investigate the exact numerical values
for the {\AA}berg criterion since our main aim is the
investigation of the interaction of a qubit with the 
TBRIM in the regimes of a thermalized quantum dot (e.g. $g \approx 15$)
or SYK black hole (e.g. $g \approx 1, V=0, J=1$).
As discussed in \cite{kolovsky2017}
the question about thermal description
of quantum chaos via effective hidden modes in the 
SYK regime remains open.

\section{Qubit interacting with TBRIM}

In order to study the decoherence of one qubit coupled to the fermionic 
system described by the TBRIM Hamiltonian $H_I$ defined in 
(\ref{eq_TBRIM}) we consider the total Hamiltonian 
\begin{equation}
\label{eq_ham_qubit}
H=\delta\cdot\,\sigma_x+
\varepsilon \frac{V_{\rm eff}}{V_0}\,\sigma_z\sum_{k=1}^{M-1} 
\left(c^\dagger_k c^\pdag_{k+1}+c^\dagger_{k+1} c^\pdag_k\right)
+H_I
\end{equation}
where $\sigma_x$ and $\sigma_z$ are the usual Pauli matrices in 
qubit space and $\delta$ is (half) the unperturbed energy separation of the 
two qubit levels introducing Rabi oscillations with 
frequency $\omega_R=2\delta$. We typically choose $\delta=\Delta_1/2$ 
(or a simple multiple of this) with $\Delta_1$ 
being the effective rescaled one-particle level spacing given above 
in terms of the effective energy scale $V_{\rm eff}$. In (\ref{eq_ham_qubit}) 
we have chosen the orbital indices $k$ such that the one-particle energies 
are ordered, i.e.~: $\epsilon_{k+1}>\epsilon_k$, implying that the 
qubit-fermion coupling term creates transitions between adjacent 
orbitals with approximate energy difference $\sim \Delta_1$. 
The quantity $\varepsilon$ is the coupling parameter which will take various 
values in the interval $0.005\le \varepsilon\le 1$ and the 
ratio $V_{\rm eff}/V_0$ (with $V_0=\sqrt{14}$) ensures that at different 
values of $V$ and $J$ the coupling parameter is measured in units of 
the overall bandwidth $\sigma\sim V_{\rm eff}$ such that results at different 
values of $V$ and $J$ at same $\varepsilon$ are indeed comparable. 
We mention that the Hamiltonian (\ref{eq_ham_qubit}) is similar in structure 
to the Hamiltonian studied in Ref. \cite{lee2005} where the qubit was coupled 
to a quantum kicked rotor model. As already mentioned we present 
below results for $M=16$ orbitals and $L=7$ particles corresponding to 
a combined qubit-fermion Hilbert space dimension of $22880$ but we have also 
verified the smaller cases at $M=12$ or $M=14$ with $L=M/2-1$
obtaining there similar results. 

Explicitly, we compute numerically the exact time evolution of a state 
$\ket{\psi_m(t)}=\exp(-iHt)\,\ket{\psi_m(0)}$ with the initial vector 
\begin{equation}
\label{eq_init_state}
\ket{\psi_m(0)}=\ket{\phi_m}(\ket{0}+2\,\ket{1})/\sqrt{5}
\end{equation}
where 
$\ket{\phi_m}$ is an exact eigenstate of $H_I$ at level number $m$ 
with many body energy $E_m$, 
i.e. $H_I\,\ket{\phi_m}=E_m\,\ket{\phi_m}$, and $\ket{0}$, $\ket{1}$ denote 
the two qubit states with bottom and upper energies. 
The time evolution operator $\exp(-iHt)$ 
is computed exactly by 
diagonalizing $H$ and expressing the matrix exponential using the exact 
eigenvalues and eigenvectors of $H$. For $M=16$ and $L=7$ 
this corresponds to a numerical diagonalization in the 
combined fermion-qubit Hilbert space of dimension $22880$. 
As in Ref. \cite{lee2005} we determine the $2\times 2$ 
density matrix $\rho_{ij}(t)$, $i,j=0,1$ from the partial trace over the 
fermionic 
states by: $\rho_{ij}(t)=\bra{i}\mbox{Tr}_{\rm ferm.} 
\left(\ket{\psi_m(t)}\bra{\psi_m(t)}\right)\,\ket{j}$. 
In absence of qubit-fermion coupling, i.e. $\varepsilon=0$, the density 
matrix $\rho(t)$ 
does not depend on the choice of $\ket{\phi_m}$ and a simple standard 
calculation gives the result:
\begin{eqnarray}
\label{eq_rho11}
\rho_{11}(t)&=&1-\rho_{00}(t)=
\frac12+\frac{3}{10}\cos(\omega_R t)\quad,\\
\label{eq_rho01}
\rho_{01}(t)&=&\rho_{10}^*(t)=\frac25+\frac{3}{10}\,i\,\sin(\omega_R t)
\quad,\\
\Rightarrow\quad |\rho_{01}(t)| &=&
\frac{1}{2}\left(\frac{41}{50}-\frac{9}{50}\cos(2\omega_R\,t)\right)^{1/2}\quad,
\end{eqnarray}
where $\omega_R=2\delta$ is the Rabi frequency. 

For practical reasons we compute the density matrix $\rho(t)$ 
at $t=\tau\,\Delta t$ with integer values of $\tau$ and the elementary time 
unit $\Delta t=1/(\Delta_1\,M)$ where 
$\Delta_1$ is the rescaled effective one-body level spacing. This time 
step corresponds roughly to the inverse one-particle band-width and represents 
the shortest quantum time scale in the system. We consider the maximal 
time value $t_{\rm max}=(d/2)\,\Delta t=5720\,\Delta t=
\sqrt{\frac{L(M-L)}{4(M-1)}}\,t_H\approx\,t_H$ with $L\approx M/2$ and 
$t_H=1/\Delta_{\rm MB}$ being the Heisenberg time. 

\begin{figure}[t]
\begin{center}
\includegraphics[width=0.48\textwidth]{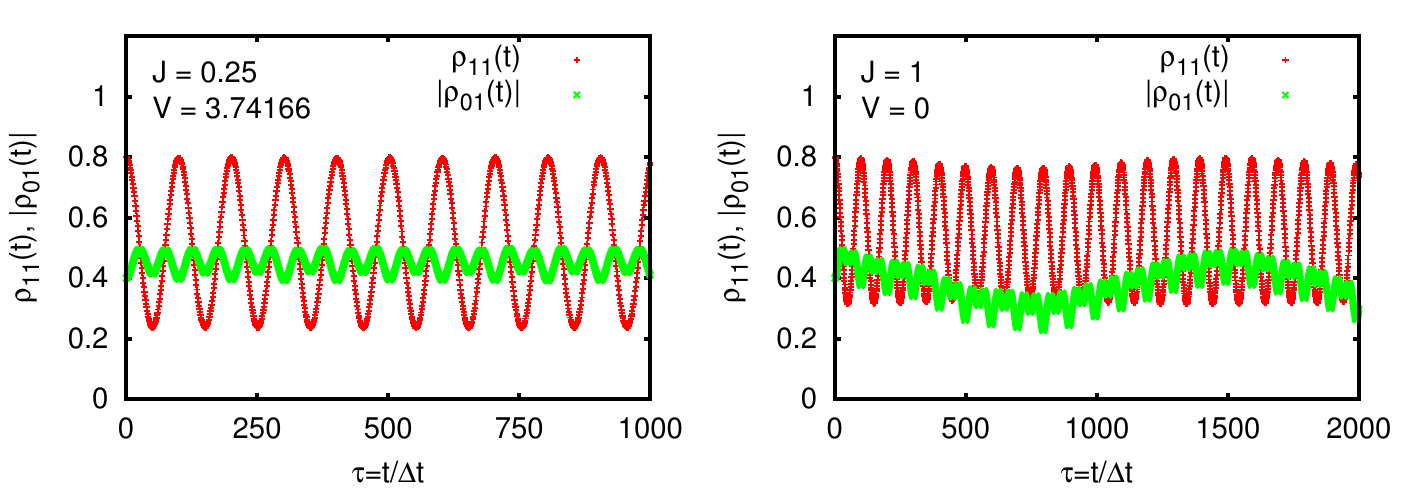}
\end{center}
\caption{\label{fig2} Time dependence of $\rho_{11}(t)$ 
(red plus symbols) and $|\rho_{01}(t)|$ (green crosses) 
for the initial state being the ground state of $H_I$ with the
level number $m=0$ and qubit state (\ref{eq_init_state})
for $V=3.74166$, $J=0.25$ ($V=0$, $J=1$) 
in left (right) panel at coupling strength $\varepsilon=0.01$. 
The time is measured in units of 
$\Delta t=1/(\Delta_1\,M)$ where $\Delta_1$ is the rescaled effective one-body 
level spacing defined in the text. }
\end{figure}

In Fig.~\ref{fig2} we show the time dependence of 
$\rho_{11}(t)$ and $|\rho_{01}(t)|$ for a weak coupling strength 
$\varepsilon=0.01$, an initial state (\ref{eq_init_state}) with 
level number $m=0$, corresponding to the ground state of $H_I$, 
and two cases for different values of $V$ and $J$.  
For $\varepsilon=0.01$ the dependence
$\rho_{11}(t)$ is close to the analytical result (\ref{eq_rho11}). 
However for $|\rho_{01}(t)|$ the situation is more 
complicated with the appearance of a further frequency leading to a 
quasi-periodic structure. Apparently the ground state $\ket{\phi_0}$ of 
$H_I$ is also weakly coupled to the next state $\ket{\phi_1}$ due to the 
indirect qubit-fermion coupling leading to an additional frequency. 
The results of Fig.~\ref{fig2} show that there is no
qubit decoherence when it is coupled with 
a quantum dot or SYK system when they are in their ground state.

\begin{figure}[t]
\begin{center}
\includegraphics[width=0.48\textwidth]{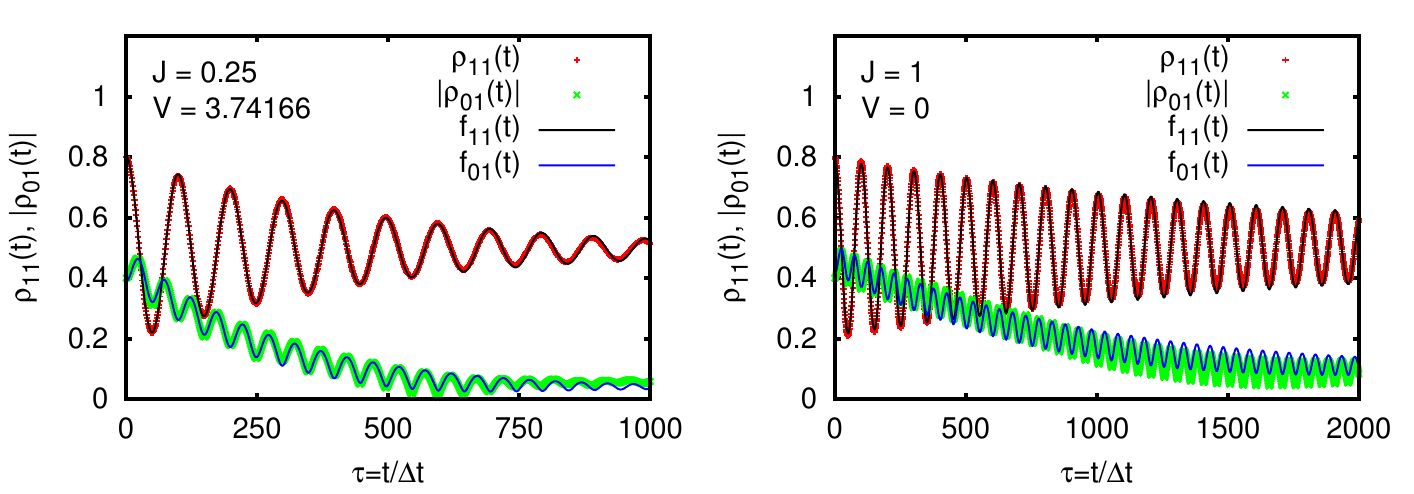}
\end{center}
\caption{\label{fig3}(color online) As in Fig.~\ref{fig2} 
but for level number $m=5720$ of the initial state (\ref{eq_init_state})
corresponding to an energy in the center of the spectrum of $H_I$. 
The fit functions $f_{11}(t)$ (thin black line) to approximate $\rho_{11}(t)$ 
and $f_{01}(t)$ (thin blue line) to approximate $|\rho_{01}(t)|$ 
are given by 
(\ref{eq_f11}) and (\ref{eq_f01}) with the fit parameters~:
$A_1=0.49593\pm 0.00005$, $B_1=0.3070\pm 0.0002$, 
$\Gamma_1=0.002195\pm 0.000003$, 
$\omega_1=0.063423\pm 0.000003$, $\alpha_1=6.2492\pm 0.0006$ and
$A_2=0.0063\pm 0.0001$, $\tilde B_2=0.194\pm 0.001$, 
$\tilde\Gamma_2=0.00435\pm0.00004$, $\omega_2=0.12641\pm 0.00004$, 
$\alpha_2=3.232\pm 0.007$, 
$B_2=0.800\pm 0.001$, $\Gamma_2=0.00713\pm 0.00002$ 
for $V=3.74166$, $J=0.25$ (left panel)
and 
$A_1=0.5008\pm 0.0001$, $B_1=0.2897\pm 0.0004$, 
$\Gamma_1=0.000449\pm 0.000002$, 
$\omega_1=0.062552\pm 0.000002$, $\alpha_1=6.228\pm 0.0002$ and
$A_2=0.0391\pm 0.0004$, $\tilde B_2=0.172\pm 0.002$, 
$\tilde\Gamma_2=0.00094\pm0.00002$, $\omega_2=0.12508\pm 0.00002$, 
$\alpha_2=3.06\pm 0.02$, 
$B_2=0.825\pm 0.002$, $\Gamma_2=0.00209\pm 0.00001$ 
for $V=0$, $J=1$ (right panel). 
}
\end{figure}

For higher level numbers $m$ the situation changes and for 
many eigenstates $\ket{\phi_m}$ of $H_I$  an exponential relaxation 
is found for $\rho_{00}(t)$ tending to 
the equilibrium value $1/2$ and $|\rho_{01}(t)|$ tending to a value 
$\sim 1/\sqrt{n}$ where $n$ is roughly the number of eigenstates of $H_I$ 
contributing in $\ket{\psi_m(t)}$. Therefore, motivated by the 
analytic expressions at $\varepsilon=0$, we use the following fit functions 
for small values $0<\varepsilon\ll 1$~:
\begin{eqnarray}
\label{eq_f11}
f_{11}(\tau\Delta t)&=&A_1+B_1\,e^{-\Gamma_1\tau}\,\cos(\omega_1\tau+\alpha_1)
\quad,\\
\nonumber
f_{01}(\tau\Delta t)&=&\frac12 \Bigl(A_2+\tilde B_2\,e^{-\tilde\Gamma_2\tau}\,
\cos(\omega_2\tau+\alpha_2)\\
\label{eq_f01}
&& +B_2\,e^{-\Gamma_2\tau}\Bigr)^{1/2}\quad,
\end{eqnarray}
to approximate $\rho_{11}(t)$ by $f_{11}(t)$ and $|\rho_{01}(t)|$ by 
$f_{01}(t)$. The parameter $\tau=t/\Delta t$ is the rescaled time in 
units of $\Delta t=1/(\Delta_1\,M)$ where $\Delta_1$ is the 
rescaled effective one-body level spacing introduced above. 
These fits work very well for the two cases shown in Fig.~\ref{fig3} with 
level number $m=5720$ (corresponding to the band center of $H_I$) and 
$\varepsilon=0.01$. 
From (\ref{eq_rho11}), (\ref{eq_rho01}) and for the choice $\delta=\Delta_1$, 
$M=16$ we expect that 
$\omega_1=\omega_R\,\Delta t=2\delta/(M\Delta_1)=1/M=0.0625$ and 
$\omega_2=2\omega_1=0.125$ which is indeed well confirmed by the 
fits shown in Fig.~\ref{fig3}. 

For larger values of the coupling strength $\varepsilon\ge 0.1$ the fits 
with the oscillatory terms do not work very well and have to be simplified 
to simple exponential fits, i.e. 
by omitting the term $\sim \tilde B_2$ in (\ref{eq_f01}) or 
replacing $\cos(\omega_1\tau+\alpha_1)\to 1$ in (\ref{eq_f11}). 
In Appendix \ref{appb} we discuss certain cases, with low values of the 
level number $m$ of the initial state (\ref{eq_init_state}) where the 
fit procedure is also problematic. However, in global the fits of the 
relaxation of the 
density matrix components work well and allow to determine the
dependence of the relaxation rates $\Gamma_1, \Gamma_2$ on system 
parameters.

\section{Qubit relaxation rates}
\label{sec4}
\subsection{Dependence on coupling strength}

The relaxation rates are computed by the methods described in 
the previous Section. Here we analyze the dependence of these rates
on system parameters. 
We note that according to usual cases of superconducting qubit relaxation
\cite{averin,shnirman,wendin,lee2005} 
the rate $\Gamma_2$ describes the dephasing of qubit
while $\Gamma_1$ describes the population relaxation.

\begin{figure}[t]
\begin{center}
\includegraphics[width=0.48\textwidth]{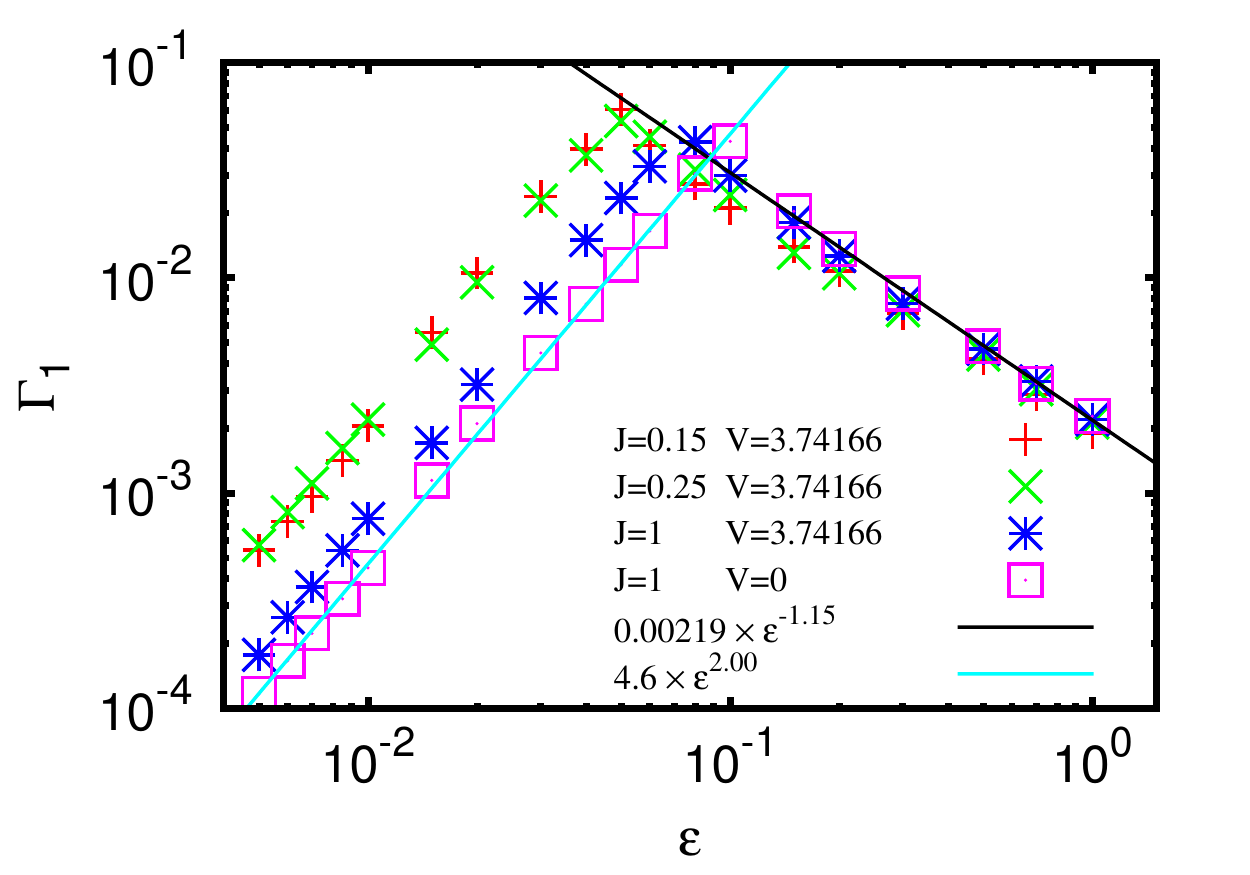}
\end{center}
\caption{\label{fig4}(color online) 
Dependence of the relaxation rate $\Gamma_1$ on the coupling
strength $\varepsilon$ at level number $m=5720$ for the initial state 
(\ref{eq_init_state}) for $V=3.74166$, $J=0.15$ (red plus symbols), 
$J=0.25$ (green crosses), $J=1$ (dark blue stars) and $V=0$, $J=1$ 
(pink squares) in a double logarithmic representation. The two lines 
correspond to the power law fits $\Gamma_1=C_1\,\varepsilon^p$ 
for $V=0$, $J=1$ with 
$C_1=4.6\pm 0.3$, $p=2.00\pm 0.02$ for $\varepsilon\le 0.1$ 
(light blue line) and $C_1=0.00219\pm 0.00006$, $p=-1.15\pm 0.02$ 
for $\varepsilon>0.1$ (black line). }
\end{figure}

The obtained dependence  of $\Gamma_1$ on 
the qubit coupling strength $\varepsilon$ 
is shown in Fig.~\ref{fig4} for the initial
state $m=5720$ taken in the middle of the total energy band
and the TBRIM values
$J=0.15,\ 0.25,\ 1$ at $V=3.74166$
($V_{\rm eff}/V_0= 1.0070$, $1.0193$, $1.2747$ and 
$\Delta_1 = 0.1475$, $0.1494$, $0.1868$ respectively) 
corresponding to the quantum
dot regime and $J=1$ at $V=0$ ($V_{\rm eff}/V_0= 0.7905,\ \Delta_1=0.1158$)
corresponding to the SYK black hole regime.
For small coupling $\varepsilon < 0.1$ the results
are well described by the quadratic dependence on coupling,
typical for the Fermi golden rule regime:
\begin{equation}
\label{eq_gammafermi1}
\Gamma_1 = C_1 \varepsilon^2 \; .
\end{equation}
The fit value of the exponent is $p=2.00 \pm 0.02$
being compatible with the quadratic dependence.

The dimensionless constant $C_1$ in (\ref{eq_gammafermi1}) is practically
independent of $J$ (at fixed $V=\sqrt{14}$)
when the system is in the regime of dynamical thermalization
being $C_1 \approx 23$ for $J=0.15,\ 0.25$ and
$C_1 \approx 8$ for $J=1$. For the SYK case we find
$C_1=4.6 \pm 0.3$ at $J=1,\ V=0$. We consider that this variation
of $C_1$ is not significant since it changes only by a factor
$5$ while $J^2$ is changed by a factor $44$ and in addition 
the model is changed from quantum dot to SYK regime.
At such changes the total energy band width is also 
changed by a factor  2 (see Fig.~\ref{fig1}) but we remind that 
due to the definition of the model and parameters in Sections 2 and 3 
both $\varepsilon$ and the relaxation rates are measured in units of 
effective energy (or inverse time) scales that take into account 
the modification of total energy band width due to different values of 
$V$ and $J$. 
We note that the dependence (\ref{eq_gammafermi1})
was also found for the dynamical relaxation
of a qubit coupled to a deterministic detector
described by the quantum Chirikov standard map \cite{lee2005}
with $C_1 \approx 0.5$ corresponding to regime of the phase 
damping noise channel \cite{chang,lee2005}. Here we obtain $C_1$
being by a factor 10 larger but in our model
(\ref{eq_ham_qubit}) the qubit is coupled with 
several TBRIM states and we assume that this is
the reason for the increase of $C_1$.

For $\varepsilon > \varepsilon_c \approx 0.1$ 
we obtain a decrease of the relaxation rate
described by the dependence
\begin{equation}
\label{eq_gammazero1}
\Gamma_1 = C_1 \varepsilon^p \; , \; p =-1.15 \pm 0.02 
\end{equation}
with $C_1 \approx 0.002$. As in \cite{lee2005}
we attribute this decrease of $\Gamma_1$ with increase of $\varepsilon$
to the quantum Zeno effect \cite{qzeno1,qzeno2}: repeated measurements
produced by a coupled detector, represented by TBRIM
in the regime of quantum chaos, reduce the relaxation rate.
In the so called ohmic relaxation regime it is
expected that $\Gamma_1 \sim {\delta}^2/\Gamma_2 
\sim B {\delta}^2/\varepsilon^2$ 
\cite{shnirman,lee2005} (here $\delta=\Delta_1/2$). 
For the model of quantum chaos detector 
it was found that $B \approx 2.7$ \cite{lee2005}.
Instead, here we find that the exponent
$|p| =1.15 \pm 0.02 \approx 1$ being significantly 
different. We attribute this difference to
the fact that in TBRIM the qubit is coupled
to many one-particle states represented by a sum 
over $k$ in (\ref{eq_ham_qubit}).
For the numerical value 
$C_1 \approx 0.002$ we find that
it is still approximately given by the relation
$C_1 \approx B {\delta}^2$ with $B \approx 0.4$
being smaller than those in \cite{lee2005}.
A surprising feature of the obtained
quantum Zeno regime is that here $\Gamma_1$ is practically independent of
parameter choice presented in Fig.~\ref{fig4}
corresponding to DTC for the quantum dot and SYK quantum chaos regimes.

The transition  between the Fermi golden mean regime
($ \Gamma_1 \propto \varepsilon^2$) and the quantum Zeno regime
($\Gamma_1 \propto 1/\varepsilon$) takes place at 
$\varepsilon_c \approx 0.07 - 1$. This corresponds to 
the relaxation rate $\Gamma_c=\Gamma_1(\varepsilon_c) \approx 0.05$
which remains practically the same for all
parameter regimes presented in Fig.~\ref{fig4}.
According to the results and arguments presented in 
\cite{lee2005,lyapunov1,lyapunov2} it is expected that 
$\Gamma_c$ is given by the Lyapunov exponent $\Lambda$
of an underlined classical dynamics of the detector coupled to 
qubit. Indeed, this was the case for the dynamical detector
considered in \cite{lee2005}, however, for the TBRIM
it is more difficult to establish what is the Lyapunov
exponent of the corresponding classical TBRIM dynamics.
It would be possible to expect that $\Gamma_c$
can be related to the Breit-Wigner width 
$\Gamma \sim J^2 \rho_c$ appearing in the TBRIM 
in the Fermi golden rule for the transition
between directly coupled states with the density
$\rho_c$ \cite{georgeot1997}. However, the independence
of $\Gamma_c$ of system parameters 
presented in Fig.~\ref{fig4} excludes this expectation.

We make the conjecture that for given parameters
$\Gamma_c$ is determined by an effective
time $T_c$ of spreading over the network of
exponentially large size 
$d$ ($\ln d \sim M$ at large $M, L$ values)
with a very small
number of links (nonzero transition matrix elements):
$N_l=K =820 \ll d = 11440$ (for $M=16$ and $L=7$). 
Such a network is similar to the small-world  networks
appearing in many cases of social relations \cite{milgram,dorogovtsev}.
It is known that a very rapid spreading takes place on such
networks for classical \cite{dorogovtsev}
and quantum spreading \cite{giraud} with a time scale
$T_c$ being only logarithmic in system size $d$
(effect of 
{\it six degrees of separation} described in \cite{milgram,dorogovtsev}).
Thus about six transitions (links) are required
to connect on average any pair of nodes on such networks 
(for the Facebook network 
there is only four degrees of separation \cite{vigna2012}).
For typical networks like Wikipedia or WWW of universities
there are only about $N_\ell \sim 10-20$ nonzero links per row/column
in the full matrix of the network of size $d \sim 10^6$ \cite{rmp2015}.

In  the TBRIM case we have a much larger number of links per row/column
and thus we expect that only about 2-4 transitions
are sufficient to connect any two nodes (levels)
of the system. 
Due to this we can expect that in this quantum small-world regime
we have $\Gamma_c \sim C_d \Delta_1 $ 
with a numerical constant $C_d \approx 0.5$.
The proportionality $\Gamma_c \propto \Delta_1$ 
appears since $\Delta_1$ plays a role of oscillator
frequency (as for an oscillator) determining the time
scale in the regime of explosive spreading over network,
$C_d$ is inversely proportional to the degree of separation of
the network which is of the order of 2-4 transitions for TBRIM
since the number of links per column is much larger than
for Wikipedia or Facebook networks.
Thus we assume that this kind of explosive spreading, 
already discussed in \cite{giraud},
is at the origin of the independence of $\Gamma_c$ of
system parameters (for the range visible in Fig.~\ref{fig4}).
We note that this kind of explosive spreading, 
with exponentially many states populated in a finite time, was also observed 
for the emergence of quantum chaos in a quantum computer core
\cite{georgeotqc2} (see e.g. Fig.6 there).

\begin{figure}[t]
\begin{center}
\includegraphics[width=0.48\textwidth]{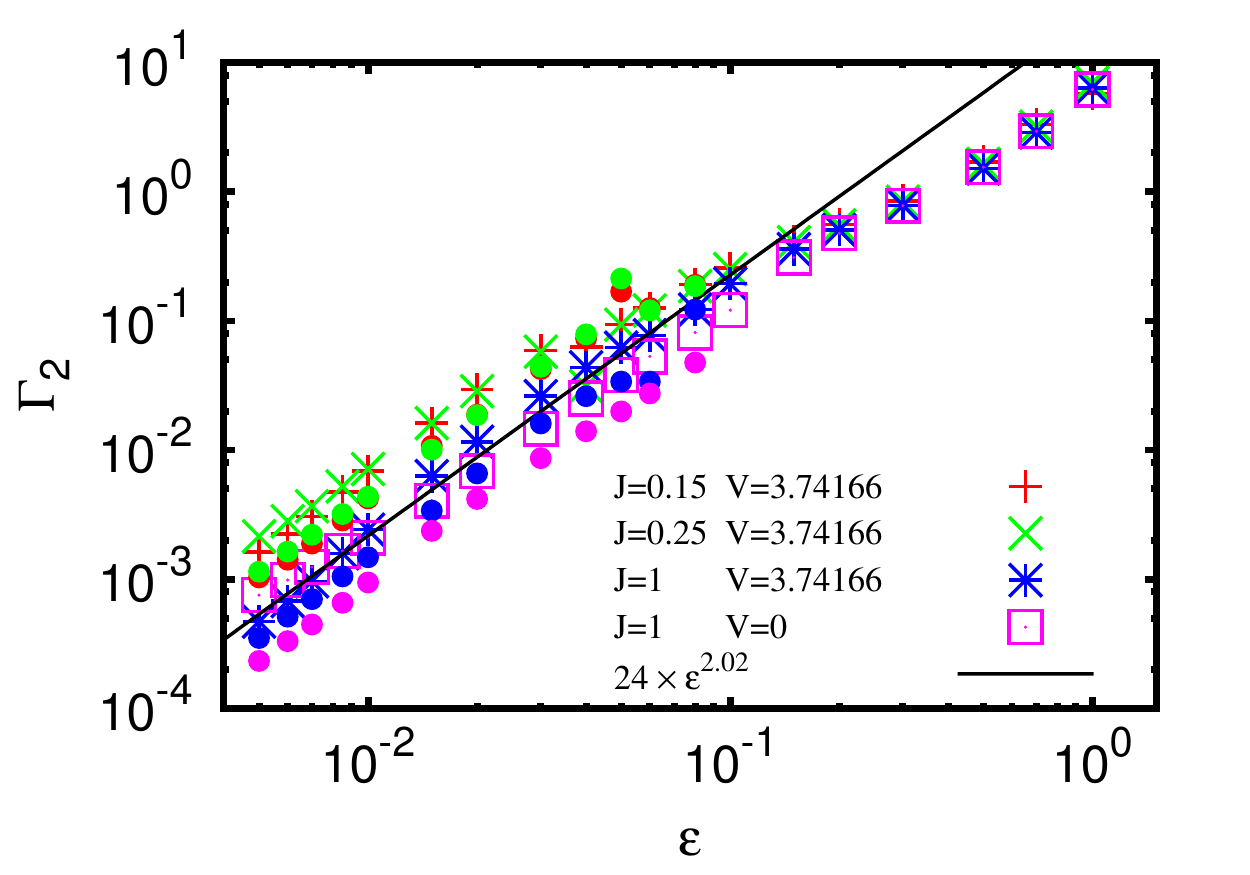}
\end{center}
\caption{\label{fig5}(color online) 
Dependence of the relaxation rate $\Gamma_2$ obtained from the 
fit (\ref{eq_f01}) on the coupling
strength $\varepsilon$ at level number $m=5720$ for the initial state 
(\ref{eq_init_state}) for $V=3.74166$, $J=0.15$ (red plus symbols), 
$J=0.25$ (green crosses), $J=1$ (dark blue stars) and $V=0$, $J=1$ 
(pink squares) in a double logarithmic representation. The black line 
corresponds to the power law fit $\Gamma_2=C_2\,\varepsilon^p$ 
for the case $V=3.74166$, $J=1$ with 
$C_2=24\pm 8$, $p=2.02\pm 0.09$ and fit range $\varepsilon\le 0.1$. 
The data points with small full circles correspond to the 
relaxation rate $\tilde\Gamma_2$ of the oscillatory term for 
$\varepsilon<0.1$ in (\ref{eq_f01}) 
(same colors as other data points for different cases 
of $V$ and $J$). For $\varepsilon\ge 0.1$ the relaxation rate 
$\Gamma_2$ is obtained from a simplified exponential fit without 
oscillatory term. }
\end{figure}

The dependence of the dephasing rate $\Gamma_2$ on the 
coupling strength $\varepsilon$ is shown in Fig.~\ref{fig5}
for the parameters considered in Fig.~\ref{fig4}.
In agreement with the usual expectations
\cite{shnirman,lee2005} we find 
\begin{equation}
\label{eq_gammazero2}
\Gamma_2 = C_2 \varepsilon^2 \; , \; C_2 = 24 \pm 8 \; . 
\end{equation}
Indeed, the numerical fit gives the exponent $p=2.02 \pm 0.09$
being very close to the Fermi golden rule value $p=2$.
For the range $\varepsilon < \varepsilon_c \approx 0.7$
we have the approximate relation $\Gamma_1 \approx \Gamma_2$
as it was also found in \cite{lee2005} 
corresponding to the general results of Ref. \cite{shnirman}.
We note that the fit results give for the other exponential decay rate 
$\tilde\Gamma_2$ of the oscillatory term in (\ref{eq_rho01}) 
(see full color circles in Fig.~\ref{fig5}) 
comparable values and parameter dependence as for $\Gamma_2$.

\subsection{Dependence on excitation level number}

\begin{figure}[t]
\begin{center}
\includegraphics[width=0.48\textwidth]{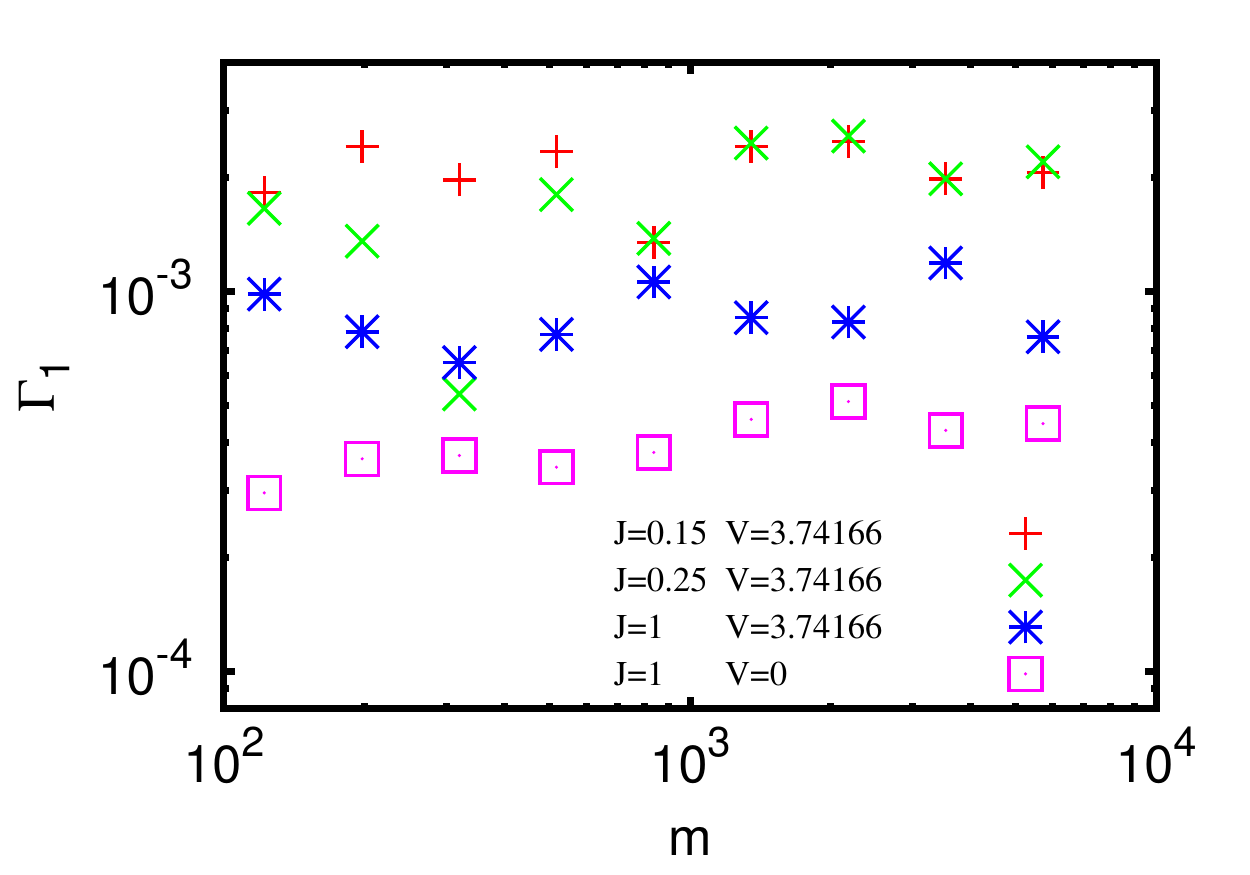}
\end{center}
\caption{\label{fig6}(color online)
Dependence of the relaxation rate $\Gamma_1$ on the level number 
$m$ used for the initial state (\ref{eq_init_state}) 
at coupling strength $\varepsilon=0.01$ for 
$V=3.74166$, $J=0.15$ (red plus symbols), 
$J=0.25$ (green crosses), $J=1$ (dark blue stars) and $V=0$, $J=1$ 
(pink squares) in a double logarithmic representation. }
\end{figure}

\begin{figure}[t]
\begin{center}
\includegraphics[width=0.48\textwidth]{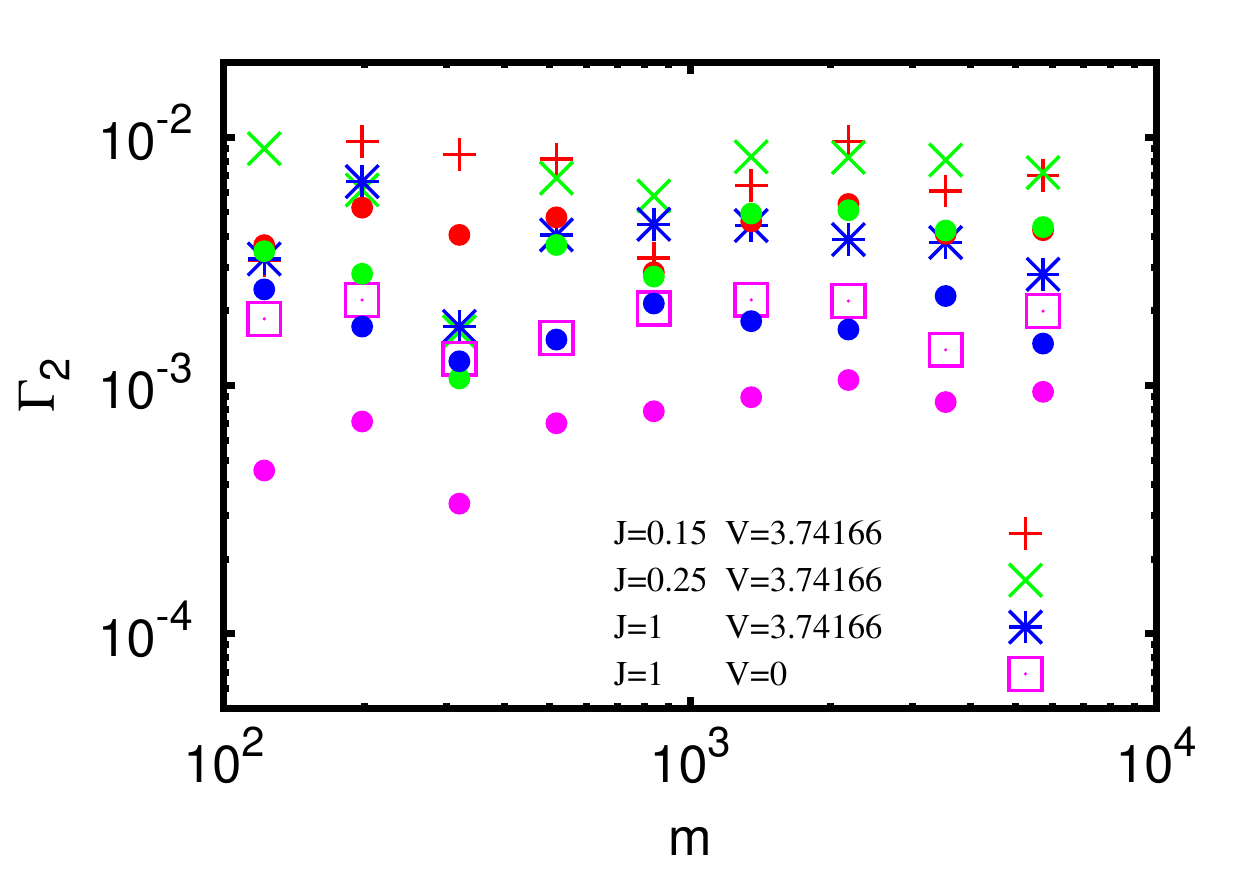}
\end{center}
\caption{\label{fig7}(color online)
Dependence of the relaxation rate $\Gamma_2$ obtained from the 
fit (\ref{eq_f01}) on the level number $m$ used for the initial 
state (\ref{eq_init_state}) at coupling strength $\varepsilon=0.01$ 
for $V=3.74166$, $J=0.15$ (red plus symbols), 
$J=0.25$ (green crosses), $J=1$ (dark blue stars) and $V=0$, $J=1$ 
(pink squares) in a double logarithmic representation. 
The data points with small full circles correspond to the 
relaxation rate $\tilde\Gamma_2$ of the oscillatory term in (\ref{eq_f01}) 
(same colors as other data points for different cases 
of $V$ and $J$).}
\end{figure}

The dependence of decay rates on the initial eigenvalue number
$m$ (with eigenstate energy  $E_{\rm ex}(m)$) is shown 
for $\Gamma_1$ in Fig.~\ref{fig6}
and $\Gamma_2$ in Fig.~\ref{fig7}. 
All data are given for a weak qubit coupling
$\varepsilon=0.01$ corresponding to the Fermi golden rule regime
in Fig.~\ref{fig4}. The independence of $m$  is surprising
since we  know that the density of coupled states for effectively
interacting electrons excited above the Fermi level $\epsilon_F$
on energy $\epsilon \approx T \ll \epsilon_F$
growth with energy as $\rho_c \approx T^3/{\Delta_1}^4$
(number of effectively interacting electrons is
$\delta n \sim T/\Delta_1$ and the effective density of
interacting two-particle states is $\rho_{2,\rm eff} \sim T/\Delta_1$ with
$\rho_c \sim \rho_{2,\rm eff} (\delta n)^2$)
and the interaction induced transition rate also 
grows with energy as $\Gamma_c \sim J^2 \rho_c 
\sim J^2 T^3/{\Delta_1}^4$
\cite{aberg1,jacquod}. Thus one could expect 
an increase of $\Gamma_1, \Gamma_2$ with an increase
of $m$. The results presented in Figs.~\ref{fig6},\ref{fig7}
clearly show no increase with $m$ for the range
$500 \leq m \leq 5720$, for the range
$100 \leq m < 500$ there is also no increase with $m$ 
but the data is more fluctuating.
These fluctuations become even stronger
for the range $0 \leq m < 100$ 
so that the fits of relaxation decay
in this range become not reliable (this is discussed in detail
in Appendix Section B).  
The increase of fluctuations at low excitation numbers $m$
is natural since for lower $m$ values we have 
a decrease of number of states
effectively coupled to the qubit. 
We note that the values of ${\tilde \Gamma_2}$, shown by full circles
in Fig.~\ref{fig7}, show a similar behavior as 
$\Gamma_2$ (with somewhat larger fluctuations
at low $m$ values since the corresponding fit
(\ref{eq_f01}) is more sensitive to errors).

Thus even if the variation of $m$ is
rather large (factor $10$ or $50$) the variation of 
$\Gamma_1, \Gamma_2$ remains in the same range 
as in Figs.~\ref{fig4},~\ref{fig5} 
being restricted approximately by a factor $5$.
We explain this independence of $m$ in 
the same manner as in previously 
arguing that for $m > 100$  the transitions between
non-interacting many-body states proceed in an explosive
spreading typical on small-world networks
in a regime $\Gamma_1 \approx \Gamma_2 \sim 30 \Delta_1 \gg \Delta_1$.
In fact for $m \approx 100, J=0.15$ we 
find $\delta E \approx T^2/\Delta_1 \approx 1.4$ with
$\Delta_1 =0.1475$ 
that gives $T \approx 0.45$  (see Fig.~\ref{fig1})
and from above estimates we obtain
$\Gamma_c/\Delta_1 \approx 30$. This ratio becomes even larger
for other parameters of Figs.~\ref{fig4},~\ref{fig6}. 

\section{TBRIM as a quantum small-world network}

\begin{figure}[t]
\begin{center}
\includegraphics[width=0.48\textwidth]{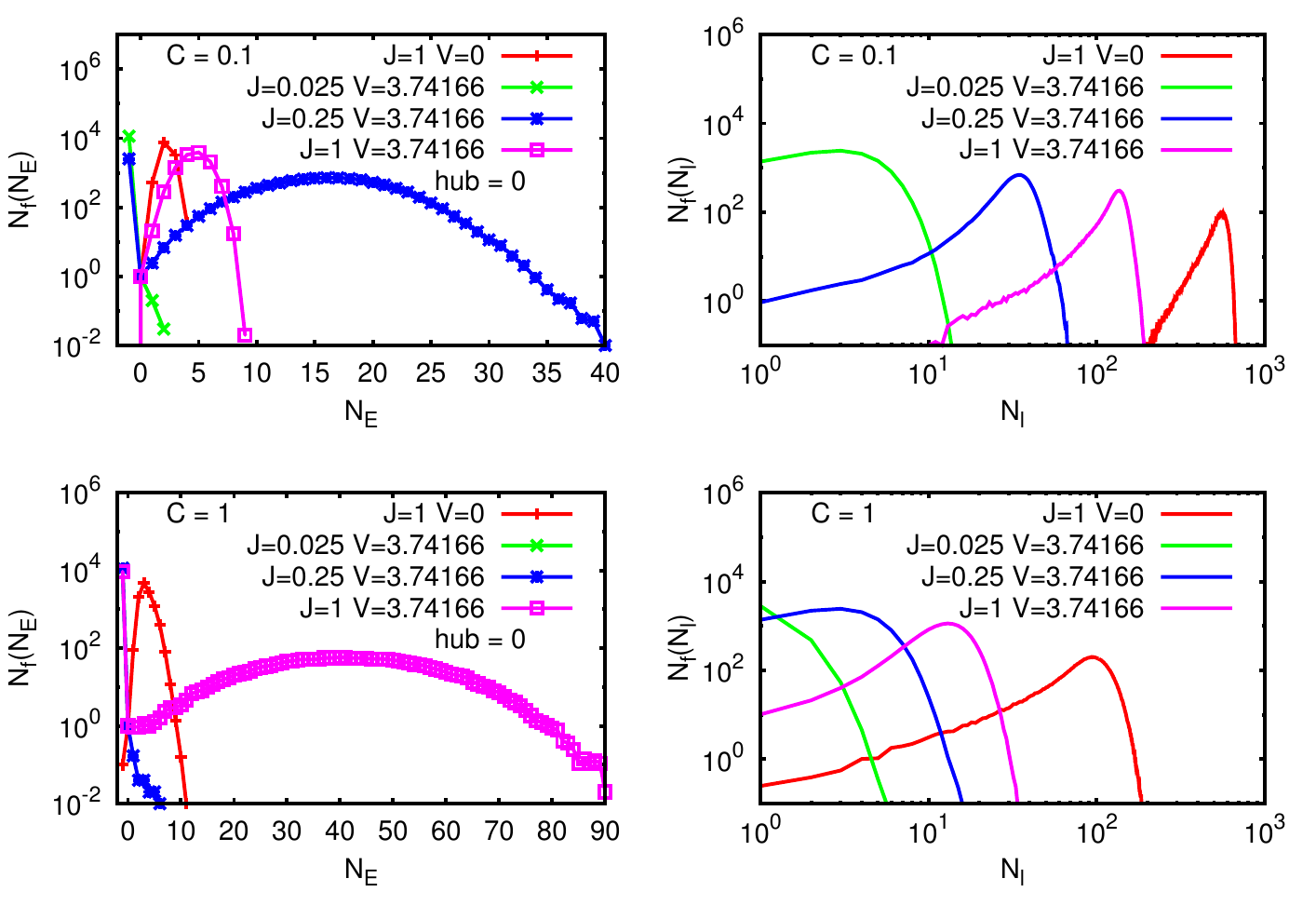}
\end{center}
\caption{\label{fig8}(color online)
Frequency distributions $N_f(N_l)$ of link number $N_l$ per node 
(right panels) and $N_f(N_E)$ of Erd\"os number $N_E$ (left panels) 
for an effective network constructed from $H_I$ where states/nodes 
$i$ and $j$ are connected by a link if the condition 
$|(H_I)_{ij}|>C|(H_I)_{ii}-(H_I)_{jj}|$ with the cut value $C=0.1$ 
(top panels) or $C=1$ (bottom panels) is met. 
The Erd\"os number $N_E$ of a node 
represents the minimal number of links necessary to connect indirectly 
this node via other intermediate nodes to the hub $=0$ 
corresponding to the many-body state where first $7$ out of $16$ orbitals 
are occupied. The hub itself has $N_E=0$ and the value $N_E=-1$ indicates 
that a node cannot be indirectly connected to the hub. Color of curves/data 
points is red (SYK, $V=0$, $J=1$), green ($V=\sqrt{14}$, $J=0.025$), blue 
($V=\sqrt{14}$, $J=0.25$) and pink ($V=\sqrt{14}$, $J=1$).
For these four cases respectively the mean and the width of the 
distribution of $N_l$ are: 
$532\pm 61$, $3.35\pm 1.87$, $33.7\pm 7.3$, $129\pm 20$ ($C=0.1$, top right 
panel) and $91.8\pm 25.2$, $0.346\pm 0.586$, $3.36\pm 1.89$, $13.4\pm 4.09$ 
($C=1$, bottom right panel); also 
the mean and the width of the distribution of $N_E$ (not counting 
$N_E=-1$ cases) are:
$2.24\pm 0.53$, $0.211\pm 0.464$, $16.1\pm 4.86$, $4.63\pm 1.12$ 
($C=0.1$, top left panel) and 
$3.40\pm 1.09$, $0\pm 0$, $0.469\pm 1.10$, $41.2\pm 13.7$ 
($C=1$, bottom left panel). 
In bottom left panel the green curve is completely hidden by the blue curve 
and contains only two values $N_f(-1)=11439$ and $N_f(0)=1$ meaning 
that the hub is not connected to any other node. 
All curves were obtained from an average of 100 different random 
realizations of $H_I$ for $M=16$, $L=7$ and $d=11440$. The vertical 
axis represents the number $N_f$ of nodes having the link number $N_l$ 
(right panels) or having the Erd\"os number $N_E$ (left panels). 
}
\end{figure}

\begin{figure}[t]
\begin{center}
\includegraphics[width=0.48\textwidth]{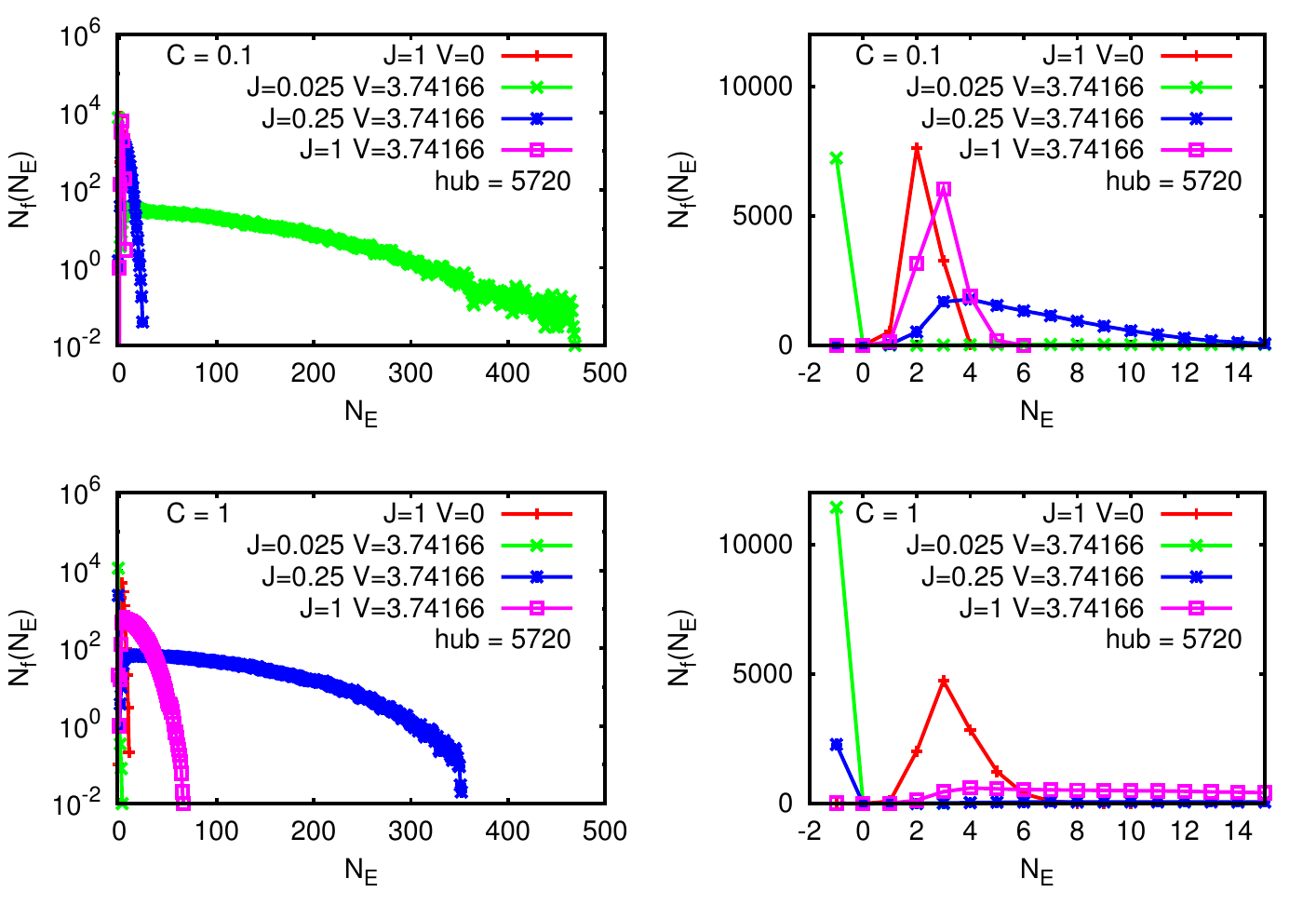}
\end{center}
\caption{\label{fig9}(color online)
Frequency distribution $N_f(N_E)$ of Erd\"os number $N_E$ 
for the same effective network of Fig. \ref{fig8} for the 
hub $=5720$ corresponding to the many-body state where orbitals:
3, 4, 8, 9, 12, 14, 15 out of $16$ orbitals are occupied. The 
signification of $N_E$ is as in Fig. \ref{fig8} and the value of 
$N_E=-1$ corresponds to the case of nodes not connected to the hub. 
Top (bottom) panels correspond to the cut value $C=0.1$ ($C=1$). 
Left panels correspond to the range $-2\le N_E\le 500$ and a logarithmic 
representation for $N_f(N_E)$ 
and right panels correspond to a zoomed range $-2\le N_E\le 15$ 
and normal representation for $N_f(N_E)$. Color of curves/data 
points is red (SYK, $V=0$, $J=1$), green ($V=\sqrt{14}$, $J=0.025$), blue 
($V=\sqrt{14}$, $J=0.25$) and pink ($V=\sqrt{14}$, $J=1$).
For these four cases respectively the mean and the width of the 
distribution of $N_E$ (not counting $N_E=-1$ cases) are:
$2.24\pm 0.53$, $96.5\pm 74.9$, $6.22\pm 3.03$, $2.90\pm 0.74$ 
($C=0.1$, top panels) and 
$3.42\pm 1.10$, $0.371\pm 0.622$, $91.7\pm 64.7$, $15.1\pm 9.5$ 
($C=1$, bottom panels). 
All curves were obtained from an average over the same 100 different random 
realizations of $H_I$ ($M=16$, $L=7$, $d=11440$) used in Fig.~\ref{fig8}. 
}
\end{figure}

Above we proposed an analogy between 
the TBRIM and a quantum small-world networks
studied in \cite{giraud,giraud2}, tracing parallels with 
the small-world networks in social relations \cite{milgram,dorogovtsev}.
On a first glance
this analogy may look to be strange since 
for TBRIM the number of nonzero
matrix elements per row/column of the Hamiltonian matrix
is fixed being $K$ while the small-world networks are 
characterized by a broader distribution
of links \cite{dorogovtsev}.
However, for the TBRIM the physical relevant quantity is not the 
formal number of nonzero elements
but the number of effectively directly coupled states.
As was discussed above and in \cite{aberg1,jacquod}
the density and number of such states depends on energy
(this is especially visible in proximity of the Fermi energy). 
According to quantum perturbation theory we need to count 
only transitions for which the transition matrix 
element is at least comparable to the energy detuning
between the states involved in the transition.
For the states with large energy detunings
the effective probabilities (weights) of the transitions
become small and their influence  can be neglected,
at least in a first approximation.

Therefore we construct from the TBRIM Hamiltonian $H_I$ defined 
in (\ref{eq_TBRIM}) an effective (symmetric undirected) network where two 
many-body states i and j are coupled by a link if the condition 
$|(H_I)_{ij}|>C|(H_I)_{ii}-(H_I)_{jj}|$ 
is met where $C$ is a parameter of order unity which we either choose 
$C=0.1$ or $C=1$. This can be considered as a numerical
selection following the {\AA}berg criterion \cite{aberg1,jacquod}.
A similar procedure has been considered in \cite{prelovsek}
for spin chains.
 We emphasize that the diagonal matrix elements $(H_I)_{ii}$ 
are constituted of two contributions: the first term in 
(\ref{eq_TBRIM}) given by the sum of energies of occupied orbitals 
and certain non-vanishing diagonal contributions from the interaction 
which have, according to the discussion in Appendix \ref{appa}, a variance 
which is $L(L-1)$ larger than the variance of the non-diagonal interaction 
matrix elements (for the case where two occupied orbitals differ between the 
two states). For the limit $J\ll V$ 
the diagonal matrix elements are of course dominated 
by the orbital energy contribution but even for the SYK case with 
vanishing orbital energies ($V=0$, $J=1$) the 
diagonal terms have a considerable 
size due to the diagonal interactions. 

Using this kind of network model we determine the frequency distribution $N_f(N_l)$
of  number of links per node $N_l$ for the four cases $V=0$, $J=1$ (SYK case with 
strongest interactions and quantum chaos), 
$V=\sqrt{14}$, $J=0.025$ (weak interactions without dynamical thermalization), 
$V=\sqrt{14}$, $J=0.25$ (moderate interactions with dynamical thermalization) 
and 
$V=\sqrt{14}$, $J=1$ (strong interactions with dynamical thermalization). 
Furthermore 
we choose our standard parameters $L=7$, $M=16$ giving a matrix dimension 
$d=11440$ of $H_I$ and the number of non-zero couplings elements per 
state $K=820$ which is an obvious upper bound for $N_l$. As can be seen 
in Fig. \ref{fig8} the frequency distribution of $N_l$ is not a power law and 
not scale free. Essentially the criterion in terms of diagonal energy 
differences implies that the typical link number $N_l$ is a certain fraction 
of $K$ which does not fluctuate too strongly for different initial states. 
However, this fraction is smallest for $V=\sqrt{14}$, $J=0.025$ with 
a maximal value $N_{f,{\rm max}}=16$ (if $C=0.1$) or $7$ (if $C=1$) 
and largest for the SYK case $V=0$, $J=1$ with 
$N_{f,{\rm max}}=687$ (if $C=0.1$) or $217$ (if $C=1$). 
According to Fig. \ref{fig8} the frequency distribution of $N_l$ 
provides largest values for the case of strongest coupling (SYK, 
$V=0$, $J=1$) and smallest values for the case of weakest coupling 
($V=\sqrt{14}$, $J=0.025$). The choice $C=1$ as compared to $C=0.1$ 
provides a general shift to smaller values. 
Actually, for $V=\sqrt{14}$ the case $C=1$, $J=0.25$ is rather comparable 
to $C=0.1$ and $J=0.025$ which is rather obvious since reducing the constant 
$C$ by a certain factor corresponds to reducing the typical interaction 
couplings by the same factor. However, these two cases are not perfectly 
identical and the remaining small differences are due to complications from 
the diagonal interaction matrix elements in $(H_I)_{ii}$. 

In global we see that the frequency distribution of links $N_f(N_l)$
is peaked near a certain average value that can be viewed as a broadening
of the delta-function distribution of random graphs introduced
by  Erd\"os-R\'enyi \cite{erdos}, known as 
the  Erd\"os-R\'enyi model \cite{dorogovtsev}. Below we check if
our quantum network possesses the small-world property
typical for the social networks \cite{milgram,vigna2012,dorogovtsev} 

With this aim we compute a more interesting quantity which we call the 
Erd\"os number $N_E$. This number represents the minimal number of links 
necessary to connect indirectly a specific node via other intermediate 
nodes to a particular node called the hub. 

We choose as hub two example states at index values $0$ and $5720$ 
in the many body Hilbert space of dimension $d=11440$ (at $M=16$ and $L=7$). 
In our numerical mapping of states (i.e. the way the many-states are 
enumerated) 
the hub $=0$ corresponds to the state where the the first $L$ of the $M$ 
(i.e. first 7 of 16) orbitals are occupied. Since we have chosen the 
orbital energies ordered with respect to the orbital index number 
(see text below Eq. (\ref{eq_ham_qubit})) 
this state corresponds to the non-interacting ground state, i.e. the Fermi sea, 
for the case $V>0$ and $J=0$. According to the Gaussian density of states 
this implies that typical energy differences of this state with the next 
excited states are rather large and therefore this hub is quite ``badly'' 
coupled to other nodes in our network model. 

The other hub $=5720$ corresponds roughly to a state in the middle of 
the non-interacting energy spectrum (for $V>0$ and $J=0$) and in our 
numerical mapping this corresponds to the state where the 7 orbitals: 
3, 4, 8, 9, 12, 14, and 15 are occupied. Here the typical energy differences 
with respect to neighbor states are quite small. 

Therefore, for the three cases with $V=\sqrt{14}$ we expect there will be 
a considerable difference in the connectivity between both hubs. However, 
for the SYK case with $V=0$ and $J=1$ the residual diagonal energies in 
$H_I$ of these states (due to the interaction) are really fully random and 
both hubs are statistically expected to be equivalent and rather well 
connected. 

The Erd\"os number corresponds roughly to the ergodic time scale (in units 
of link-iterations) for the classical stochastic dynamics induced by 
the network. Depending on the typical coupling strength of the network 
it is possible that certain or even many nodes are not at all coupled to the 
hub by indirect links, especially for the hub $=0$ (if $V>0$).
In this case we attribute artificially the 
value $N_E=-1$ to such topologically separated nodes from the hub, 
while the hub itself has $N_E=0$ and the remaining nodes 
(indirectly coupled to the hub) have values $N_E>0$. 

The frequency distribution $N_f(N_E)$ of the Erd\"os number $N_E$ 
for hub $=0$ is shown in the left panels of Fig.~\ref{fig8}. 
For the SYK case (strongest coupling) the distribution is strongly peaked 
with typical values at $\sim 2$ ($\sim 3$) for $C=0.1$ ($C=1$). 
Then with decreasing coupling (or increasing value of $C$) 
the width and mean values of the distribution 
increase provided there is still a sufficient fraction of nodes (indirectly) 
coupled to the hub. For the cases of weakest coupling $V=\sqrt{14}$, $J=0.025$ 
(if $C=0.1$) or $J\le 0.25$ (if $C=1$) nearly all nodes are not at all 
coupled to the hub as can be seen from the strong peaks at $N_E=-1$. 
The mean and width of the distribution of the few number of remaining nodes 
(eventually only the hub itself) is very small. 
For the two cases $V=\sqrt{14}$, $J=0.25$ (if $C=0.1$) or $J=1$ (if $C=1$) 
there is a large fraction of isolated nodes but there are still enough 
remaining nodes coupled to the hub providing a non-trivial distribution 
of largest values $\sim 40$ or $\sim 90$ respectively. 
Apart from the SYK cases only the case $V=\sqrt{14}$, $J=1$ at $C=0.1$ 
provides a strongly peaked distribution with typical value at $4.6\pm 1$. 

Fig.~\ref{fig9} shows the frequency distribution $N_f(N_E)$ of the 
Erd\"os number $N_E$ for the other hub $=5720$. As expected the two 
SYK cases are very similar to the first hub $=0$ of Fig.~\ref{fig8}. 
However for the cases with $V=\sqrt{14}$ the connectivity is indeed ``better'' 
as compared to Fig.~\ref{fig8}, i.e. either the typical values are smaller 
or there are less isolated nodes (lower or absent peaks at $N_E=-1$). 
Especially the two cases $J=0.025$ (if $C=0.1$) or $J=0.25$ (if $C=0$), 
with nearly only isolated nodes in Fig.~\ref{fig8}, provide now 
a non-trivial rather large distribution for a modest fraction of non isolated 
nodes. Furthermore, these two cases are actually quite comparable as 
already discussed above for the frequency distribution of links.

\begin{figure}[t]
\begin{center}
\includegraphics[width=0.48\textwidth]{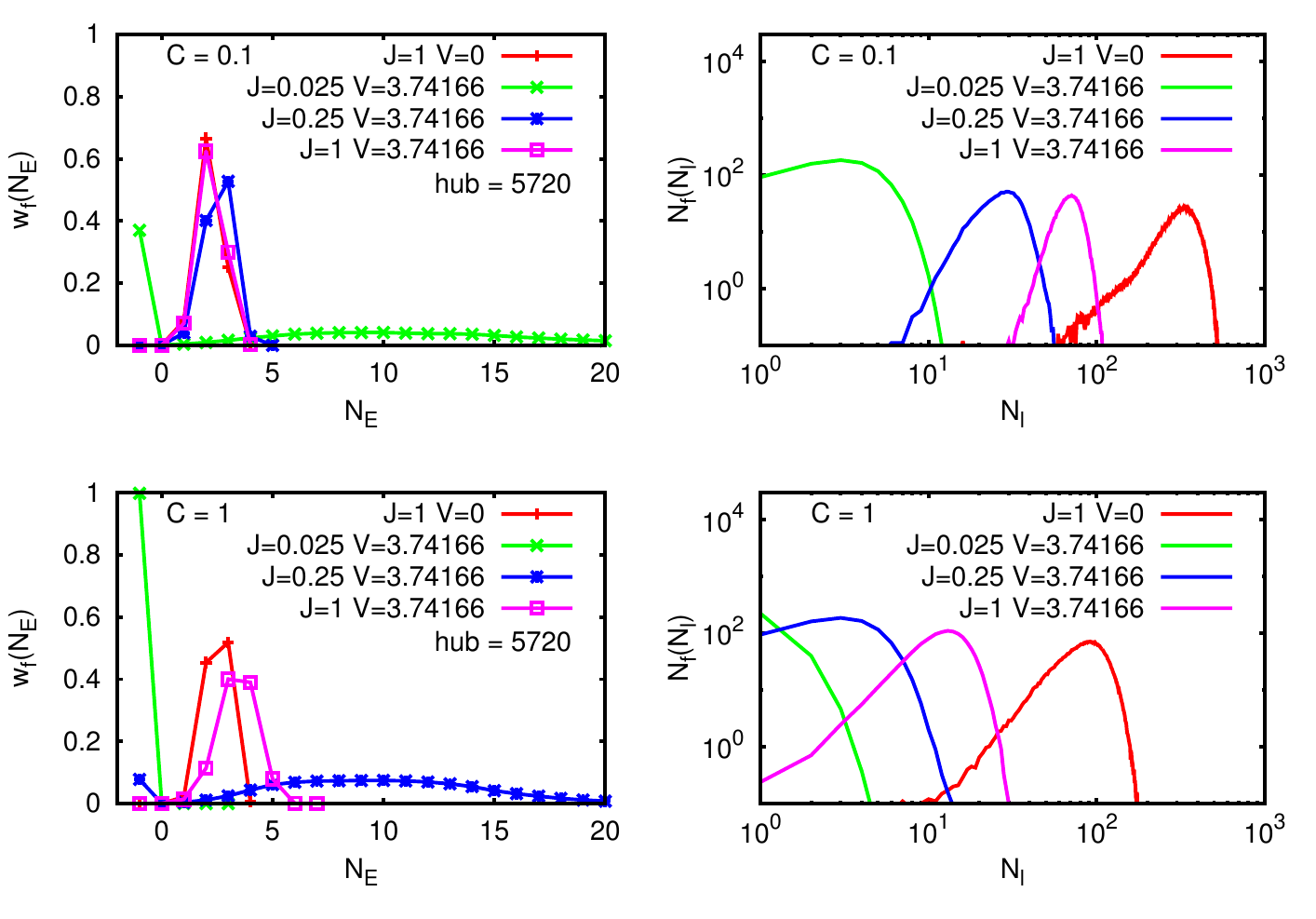}
\end{center}
\caption{\label{fig10}(color online)
Frequency distribution $N_f(N_l)$ of link number $N_l$ per node 
(right panels) and probability distribution $w_f(N_E)$ of 
Erd\"os number $N_E$ (left panels) 
for an effective network constructed from $H_I$ as in Fig.~\ref{fig8} 
but only using nodes/states satisfying the energy condition~: 
$|(H_I)_{ii}-(H_I)_{{\rm hub,hub}}|<1.5\Delta_1$ for hub $=5720$. 
Color of curves/data 
points is red (SYK, $V=0$, $J=1$), green ($V=\sqrt{14}$, $J=0.025$), blue 
($V=\sqrt{14}$, $J=0.25$) and pink ($V=\sqrt{14}$, $J=1$).
For these four cases respectively the mean and the width of the 
distribution of $N_l$ are: 
$323\pm 67$, $3.50\pm 1.87$, $28.8\pm 7.0$, $69.9\pm 10.4$ ($C=0.1$, top right 
panel) and $89.3\pm 23.4$, $0.368\pm 0.603$, $3.50\pm 1.90$, $13.4\pm 4.0$ 
($C=1$, bottom right panel); 
the mean and the width of the distribution of $N_E$ (not counting 
$N_E=-1$ cases) are:
$2.18\pm 0.56$, $12.0\pm 6.1$, $2.54\pm 0.63$, $2.23\pm 0.58$ 
($C=0.1$, top left panel) and 
$2.51\pm 0.55$, $0.371\pm 0.622$, $10.2\pm 4.5$, $3.41\pm 0.85$ 
($C=1$, bottom left panel); the average 
effective dimension/reduced network size 
is: $4107$, $869$, $883$, $1102$ (all panels). 
Left panels show the probability distribution $w_f(N_E)$ normalized to unity 
for a better visibility as compared to $N_f(N_E)$ (shown in Figs.~\ref{fig8} 
and \ref{fig9}) with different 
normalizations due to different network sizes. 
As in Fig.~\ref{fig8} the case $N_E=-1$ represents nodes 
which cannot be reached by the hub. 
All curves were obtained from an average over the same 100 different random 
realizations of $H_I$ ($M=16$, $L=7$, $d=11440$) used in Fig.~\ref{fig8}. 
}
\end{figure}

The last two cases correspond to a (partial) ergodicity but only after 
a large number of network iterations. This observation may be related to a 
diffusive dynamics in energy space where it takes some time to explore 
different energy layers such that the networks are not really of 
small-world type. Therefore we also consider a reduced network 
where we keep only nodes/states whose diagonal energies are relatively close 
to the diagonal energy of the hub, i.e. such that the energy condition 
$|(H_I)_{ii}-(H_I)_{{\rm hub,hub}}|<1.5\Delta_1$ for the hub $=5720$ 
is satisfied and where $\Delta_1$ is the 
effective rescaled average one-particle level spacing introduced 
in Section 2 (see text below Eq. (\ref{eq_a_coeff})). We remind that 
$\Delta_1$ is small compared to the overall energy band width but 
typically large compared to the many body level spacing and also with 
respect to the effective level spacing of directly interaction coupled states 
\cite{jacquod,georgeot1997}. 
As a consequence of this condition the effective dimension 
or network size of remaining nodes/states is considerably reduced to 
values $\sim 4000$ for the SYK case or $\sim 1000$ for the three 
cases with $V=\sqrt{14}$. 
The modified distributions for this reduced network 
of link number $N_l$ and Erd\"os number $N_E$ 
are shown in Fig.~\ref{fig10}. The frequency distribution $N_f(N_l)$ 
is similar as in Fig.~\ref{fig8} with a clear ordering of typical sizes 
from strongest coupling (SYK) to weakest coupling ($V=\sqrt{14}$ and 
$J=0.025$) and an overall shift from $C=0.1$ to $C=1$. 
The distribution of Erd\"os numbers for SYK is not changed (apart 
from the modified normalization) while the cases with $V=\sqrt{14}$ are 
now generally closer to a small-world situation. Here $J=1$ is now 
identical (close) to SYK, $J=0.25$ provides 
typical Erd\"os numbers $\sim 2-3$ ($\sim 10$), and the 
case $J=0.025$ corresponds to a typical Erd\"os number $\sim 12$ 
(majority of nodes isolated from hub) all for $C=0.1$ ($C=1$). This clearly 
confirms that large Erd\"os numbers $\sim 10^2$ of the full network before 
correspond to diffusion to other energy layers. 

The data of Figs.~\ref{fig8}-\ref{fig10} 
clearly show that the TBRIM is characterized by
small-world properties provided the interaction strength is sufficiently 
large. Especially the SYK case with an average Erd\"os number 
$\langle N_E\rangle =2.2\pm 0.5$ ($3.4\pm 1$) for $C=0.1$ ($C=1$) 
shows very strong small-world properties. 
However, for modest interaction strength there are some complications 
due to diffusion in energy space leading to possible 
Erd\"os numbers $\sim 10^2$. 
We think that the further development of the analogy between
quantum many-body interacting systems and small-world networks
will bring a better understanding of these quantum systems.

We note that the small-world network constructed for an energy layer 
of a finite width (we use the width of $3\Delta_1$) is more relevant for the 
qubit relaxation analyzed in previous Sections: the coupling of the qubit
with the states inside this layer leads to its rapid relaxation on
the time scale related to $\Gamma_c \sim \Delta_1$, while  slow transitions
from one energy layer to another layer describe the residual level of 
density matrix relaxation analyzed in the next Section.

\section{Residual level of density matrix relaxation}

Our TBRIM model contains a finite number of states $d$
and hence the relaxation of density matrix components
stops at a certain residual level of density matrix elements
$|\rho_{01}|$ determined by quantum deterministic fluctuations and noise.
In fact since the spectrum of our system is discrete
and the system is bounded we will always have
the Poincar\'e recurrences to the initial state
in agreement with the  Poincar\'e recurrence theorem \cite{poincare}.
However, the time $t_r$ of such a recurrence grows exponentially
with the system size $\ln t_r \propto d$ being enormously 
large even for our case with $L=7$ particles.
However, depending on the initial state and parameters it is possible 
that the effective number $d_{\rm eff}$ of excited states contributing in the 
exact time evolution is much smaller than $d$ implying that for these cases 
$t_r$ is strongly reduced. 
Therefore we compute the deterministic residual level of quantum fluctuations
given by $|\rho_{01}|$ averaged over long times for  
$d/4\le t/\Delta t\le d/2$ roughly corresponding to 
$t_H/2\le t \le t_H$ (or $|\rho_{00}-1/2|$ with similar results).

\begin{figure}[t]
\begin{center}
\includegraphics[width=0.48\textwidth]{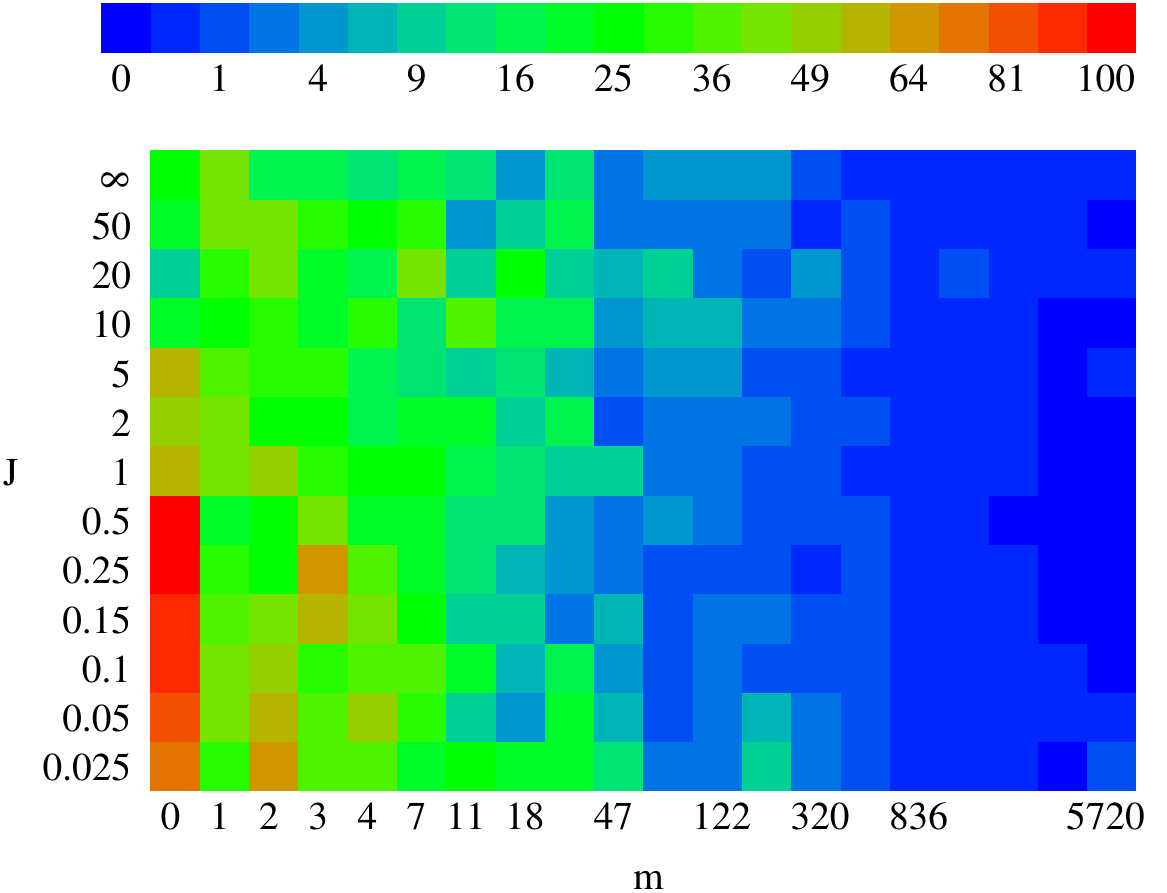}
\end{center}
\caption{\label{fig11}(color online)
Density plot of residual level of quantum fluctuations determined as the time average of 
$|\rho_{01}(t)|$ at long times for $d/4\le /\Delta t\le d/2$.
The horizontal axis corresponds to the level number 
$m=0,1,2,3,4,7,11,18,29,$ $47,76,122,198,320,517,836,1353,2187,3537,5720$
of the initial state (\ref{eq_init_state}) and the vertical axis corresponds 
to the value of the interaction strength $J$ at $V=\sqrt{14}\approx 3.74166$
except for the top row (with symbol ``$J=\infty$'') representing the 
SYK case $J=1$ and $V=0$. The coupling strength is $\varepsilon=0.03$. 
The colors red, green or blue correspond to maximum $|\rho_{01}(t)|=0.4353$, 
intermediate or minimum (zero) fluctuation values
(they are shown by color bar on top with numbers showing the percentage 
of maximal value).
}
\end{figure}

The dependence of the residual level of quantum fluctuations
on $J, m$ is shown in Fig.~\ref{fig11}. The lowest level
is at the middle of energy band with $m=5720$
corresponding to infinite temperature $T$,
 The highest level is found for the ground state $m=0$ 
and first excited states $m=1,2$ with $J<1$ at $V=\sqrt{14}$.
The amplitude of residual fluctuations
decreases with increase of $J$ but it is difficult to
establish a clear border in $(J,m)$ plane.
We attribute this to the fact that the {\AA}berg border
(\ref{eq:abergcriterion}) works mainly for small $J$ values
with $g \gg 1$ so that a special analysis of this region
is required that was not the main aim of this work.

We note that the residual fluctuations are rather similar 
for the SYK regime at $J>10$ and the quantum dot regime
above the   {\AA}berg border (\ref{eq:abergcriterion})
with $0.15 \leq J < 10$ (except very low excited levels
$m<7$ and $J<0.5$). We attribute this
to the fact that in this region
$\Gamma_c \gg \Delta_1$ leading to the explosive spreading
over the quantum small-world.

In analogy with \cite{lee2005} we expect that 
in the regime of developed quantum chaos
the residual level $R_q$
of quantum fluctuations of qubit drops 
as a square-root of the states of a detector 
$R_q \propto 1/\sqrt{d}$. However, the quantum computations
for TBRIM detector are more complicated
compared to the kicked rotator case
and we did not performed detailed numerical 
checks of this relation which is 
however in a qualitative agreement
with the results of Fig.~\ref{fig11}.

\begin{figure}[t]
\begin{center}
\includegraphics[width=0.48\textwidth]{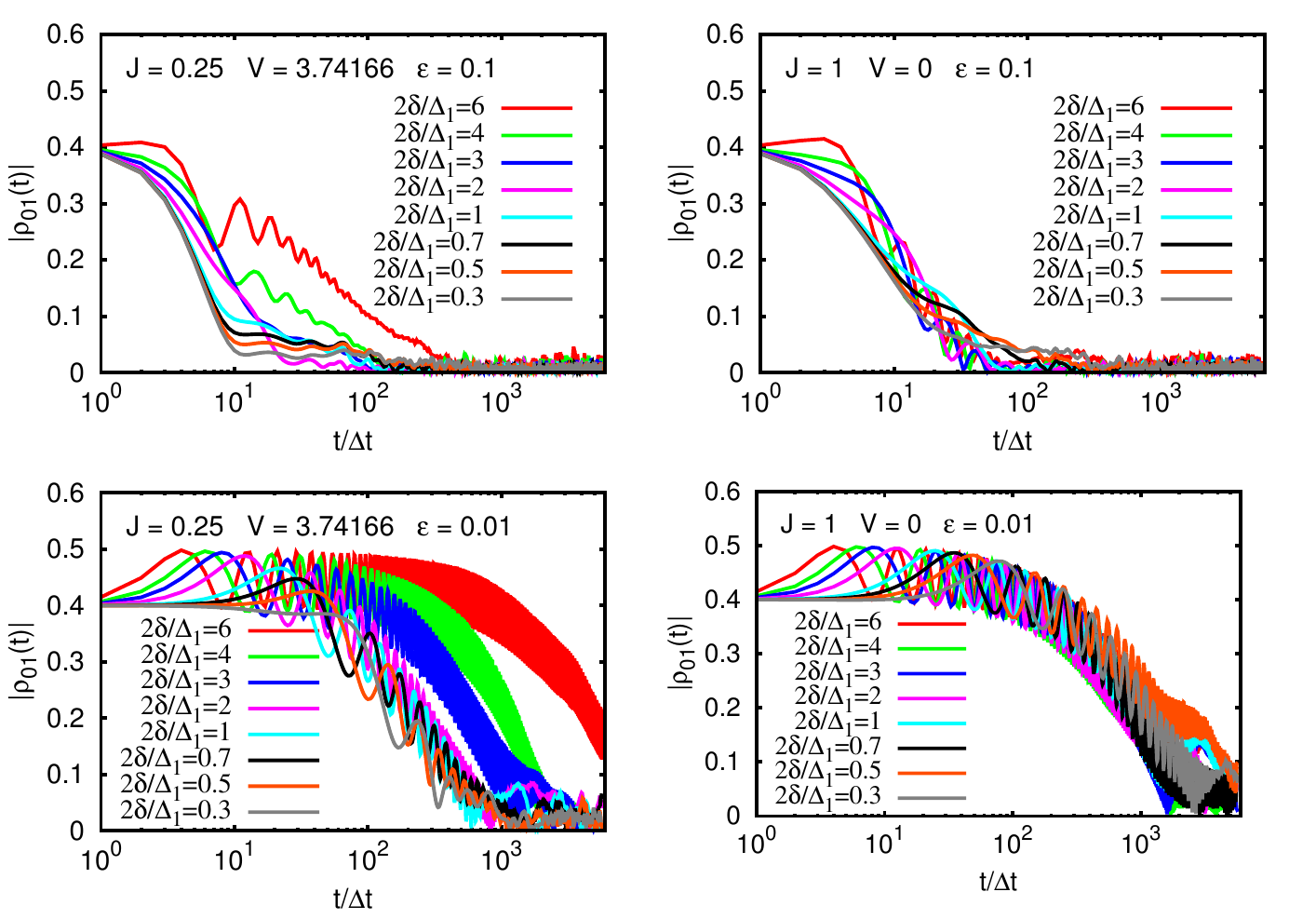}
\end{center}
\caption{\label{fig12}(color online)
Time dependence of $|\rho_{01}(t)|$ at level number $m=5720$ for 
the initial state in (\ref{eq_init_state}) for various values of 
the parameter $\delta$ according to: $0.3\le 2\delta/\Delta_1\le 6$. 
The horizontal axis for the time is shown in logarithmic representation 
for a better visibility. Top (bottom) panels correspond to coupling 
strength $\varepsilon=0.1$ ($0.01$). Left (right) panels correspond 
to $V=3.74166$, $J=1$ ($V=0$, $J=1$). 
}
\end{figure}

Finally we make a note on the relaxation dependence of the qubit energy 
given by $2\delta$. Above we presented results
for a fixed value $\delta = \Delta_1$ but we checked that
the relaxation of density matrix components
goes in a similar manner for other values of the ratio
$0.3 \leq 2\delta/\Delta_1 <3$ as it is shown in Fig.~\ref{fig12}
The changes of the decay curves start to be visible 
for $2\delta/\Delta_1 \ge 3$ but in this range the qubit
energy becomes comparable to the energy size of the TBRIM band
that corresponds to another physical regime 
where the qubit cannot be considered as a weak perturbation.

We also mention that our above discussion of 
the properties of qubit relaxation concern the range of positive 
temperatures with $m \leq d/2$.
The regime of negative temperatures is briefly discussed in
Appendix C where we find comparable results for the qubit relaxation as 
in the regime of positive temperatures. 
This is also in agreement with spin relaxation at 
negative temperatures considered in \cite{abragam}.

\section{Discussion}

We presented results for a dynamical decoherence 
of a qubit weakly coupled to the TBRIM system
in the regime of dynamical thermalization induced
by interactions and quantum many-body chaos,
corresponding to the quantum dot of interacting fermions
and the SYK black hole model. The relaxation rates of qubit population $\Gamma_1$ and 
dephasing $\Gamma_2$ are determined as a function of qubit coupling 
strength $\varepsilon$  with $\Gamma_1 \propto \varepsilon^2$
in the Fermi golden rule regime and  $\Gamma_1 \propto 1/\varepsilon$
in the quantum Zeno regime with $\Gamma_2 \propto \varepsilon^2$
for the whole considered range. These results are in a satisfactory
agreement with the usual thermal bath qubit decoherence 
considered in the literature (see e.g. \cite{shnirman}).
The surprising finding of our studies is that
the values of  $\Gamma_1, \Gamma_2$ remain practically unchanged 
in a broad range of
parameters of the quantum dot or the SYK model.
We propose a tentative explanation of this effect by 
tracing an analogy between TBRIM system
and quantum small-wold networks with appearance
of explosive spreading over exponential number of sites
(states) in a finite time. This explosive spreading appears
in both regimes of quantum dot and SYK when the
transition rates between directly coupled states become larger 
than an effective level spacing between one-particle states.
We hope that our results will stimulate further
investigations of dynamical decoherence in quantum many-body 
interacting systems and a further development of parallels between
these systems and the small-world networks.

 This work was supported in part by the Pogramme Investissements
d'Avenir ANR-11-IDEX-0002-02, reference ANR-10-LABX-0037-NEXT 
(project THETRACOM);
it was granted access to the HPC resources of 
CALMIP (Toulouse) under the allocation 2017-P0110.

\appendix
\bigskip
{\bf APPENDIX}

\section{Gaussian density of states}

\label{appa}
\subsection{Analytical computation of the variance}

The TBRIM Hamiltonian $H_I$ given by (\ref{eq_TBRIM}) exhibits 
in the limit $M\to\infty$ at a fixed value of particle number $L$ 
an average density of states which is obviously Gaussian in absence of 
interaction ($J=0$) since in this case the many body energy levels are given 
by $E(\{n_j\})=\sum_j n_j\,\epsilon_j$ with $n_j\in\{0,\,1\}$ which is a 
sum of random Gaussian variables with vanishing average and variance:
\begin{eqnarray}
\nonumber
\sigma_\epsilon^2&=&\left\langle E(\{n_j\})^2\right\rangle 
= \sum_{j,j'=1}^M n_j\,n_{j'}\,
\langle \epsilon_j\,\epsilon_{j'}\rangle\\
\label{eq_envar1}
&=&\frac{1}{M}\sum_{j=1}^M n_j^2 V^2=\frac{L}{M}\,V^2\ .
\end{eqnarray}
However, the expression (\ref{eq_envar1}) requires to take the ensemble 
average over the one-particle energies $\epsilon_j$, i.e. the numerical 
verification of the variance requires an average over many realizations and 
from a pure mathematical point of view the Gaussian form of the distribution 
of $E(\{n_j\})$ requires indeed the limit $M\to\infty$ at fixed value $L$ 
(and of $V^2/M\to$const.) providing a sum of {\em independent} Gaussian 
variables. 

On the other hand, we find numerically that the density of states is 
very close to a Gaussian distribution already for one sample of $H_I$ at the 
values of $M$ and $L$ we considered. To understand this let us first 
consider $J=0$ and let $\epsilon_j$ be 
one sample of one-particle energies initially drawn from a 
Gaussian distribution 
(with zero mean and variance $V^2/M$) and then slightly modified by a small 
universal shift and rescaling factor to ensure {\em exactly} that 
$\sum_j \epsilon_j=0$ and $\sum_j \epsilon_j^2=V^2$. Now we consider this 
set of one-particle energies fixed and perform 
the average of $E(\{n_j\})$ not with respect to $\epsilon_j$ 
but with respect to all configurations $n_j\in\{0,\,1\}$ such that 
$L=\sum_j n_j$. In this case we have obviously $\langle n_j\rangle =L/M$. 
Furthermore we find:
\begin{equation}
\label{eq_L2average}
L^2=\sum_{j,j'=1}^M \langle n_j\,n_{j'}\rangle=
M\langle n_j\rangle +M(M-1) \langle n_j\,n_{j'}\rangle_{j\neq j'}
\end{equation}
where we have separated the terms with $j=j'$ from those with $j\neq j'$. 
From (\ref{eq_L2average}) we immediately find:
\begin{equation}
\label{eq_njnjp}
\langle n_j\,n_{j'}\rangle_{j\neq j'}=\frac{L(L-1)}{M(M-1)}
\end{equation}
and therefore we get, for $J=0$, a different variance with respect to 
(\ref{eq_envar1}): 
\begin{eqnarray}
\nonumber
\sigma^2(0)&=&\left\langle E(\{n_j\})^2\right\rangle = 
\frac{L}{M}\sum_{j=1}^M \epsilon_j^2
+\frac{L(L-1)}{M(M-1)}\sum_{j\neq j'}^M \epsilon_j\,\epsilon_{j'}\\
\label{eq_envar2}
&=&\frac{L}{M}\left(1-\frac{L-1}{M-1}\right)\sum_{j=1}^M \epsilon_j^2
=\frac{L(M-L)}{M(M-1)}\,V^2\ .
\end{eqnarray}
To obtain (\ref{eq_envar2}) we have used that:
\begin{equation}
\label{eq_used}
\sum_{j\neq j'}^M \epsilon_j\,\epsilon_{j'}=\left(
\sum_{j=1}^M \epsilon_j\right)^2-\sum_{j=1}^M \epsilon_j^2
=-\sum_{j=1}^M \epsilon_j^2
\end{equation}
since $\sum_j\epsilon_j=0$ by choice. 

Now we consider a non-vanishing interaction strength $J\neq 0$. 
In the limit for sufficiently small $J$ we expect that the density of 
states is not affected by $J$. If we assume that the density of states 
remains Gaussian, also for larger values of $J$, 
(see below for the numerical confirmation of this) we 
can compute the variance $\sigma^2$ from the average:
\begin{eqnarray}
\nonumber
\sigma^2&=&\frac{1}{d}\int_{-\infty}^\infty E^2\,\langle\rho(E)\rangle\,dE
=\frac{1}{d}\left\langle \sum_{m=0}^{d-1} E_m^2\right\rangle\\
\label{eq_envar3}
&=& \frac{1}{d} \langle\mbox{Tr}(H_I^2)\rangle
=\sigma^2(0)+\frac{1}{d}\langle\mbox{Tr}(H_J^2)\rangle
\end{eqnarray}
where $E_m$ are the exact many body energies, 
$\sigma^2(0)$ is the variance at $J=0$ given in (\ref{eq_envar2}) and 
\begin{equation}
\label{eq_HJ}
H_J=\frac{4}{\sqrt{2M^3}}\sum_{i<j,k<l} J_{ij,kl}\,c^\dagger_i c^\dagger_j 
c^\pdag_l c^\pdag_k
\end{equation}
is the interaction contribution in (\ref{eq_TBRIM}). 
In (\ref{eq_envar3}) the average is done at fixed 
one-particle energies with respect to the different configurations of 
the occupation numbers $n_j$ 
(satisfying $\sum_j n_j=L$) and with respect to the Gaussian interaction 
matrix elements $J_{ij,kl}$. To evaluate $(1/d)\langle\mbox{Tr}(H_J^2)\rangle$ 
let us consider one particular many body state where exactly 
$L$ of the $M$ orbitals are occupied. 
This state is coupled by the interaction to three groups of other states: 
(i) ``itself'', i.e. with identical occupation numbers $n_j$, (ii) $L(M-L)$ 
states that differ exactly for one particle occupying another orbital, and 
(iii) $L(L-1)(M-L)(M-L-1)/4$ states that differ exactly for two particles 
occupying other orbitals. This corresponds to a total number of coupled states 
$1+L(M-L)+L(L-1)(M-L)(M-L-1)/4$, an expression already given 
in \cite{jacquod,flambaum}.

However, in 
order to evaluate the contributions of the corresponding interaction 
matrix elements in $\langle\mbox{Tr}(H_j^2)\rangle$ 
this (global) number is not relevant since the average 
variance of the interaction matrix element differs between these three 
groups. The interaction matrix element of the state with itself, 
corresponding to the group (i), uses 
$L(L-1)/2$ terms of (\ref{eq_HJ}) since there are $L(L-1)/2$ possibilities 
to destroy a pair of particles in the set of given $L$ particles 
and to recreate them afterwards in their same original orbitals. This 
corresponds to a sum of $L(L-1)/2$ independent Gaussian variables 
$J_{ij,ij}$ with variance\footnote{It is mathematically also possible 
to consider other 
symmetry classes GUE or GSE for the interaction matrix which would imply 
a variance of $2J^2/\beta$ (with $\beta=1$ for GOE, $2$ for GUE and 
$4$ for GSE) for the variables $J_{ij,ij}$
if we keep the non-diagonal variance $J^2$ of $J_{ij,kl}$ for 
$(ij)\neq (kl)$.} $2J^2$, 
giving a contribution in $\langle\mbox{Tr}(H_j^2)\rangle$ being 
$(8/M^3)\,J^2 L(L-1)$. 

Concerning the group (ii), we need to consider in (\ref{eq_HJ}) the 
index pairs $i<j$ and $k<l$ where one index of the first pair 
is identical to one index of the other pair and the other one is 
different. This gives $L-1$ possibilities to destroy the pair of particles 
and recreate them afterwards such that one of the two particles stays 
in the same orbital and the other one has changed its orbital. 
Therefore the total contribution of all states of the group (ii) to 
$\langle\mbox{Tr}(H_j^2)\rangle$ is $(8/M^3)\,J^2 L(M-L)(L-1)$. 

Concerning the group (iii) both indices must be different and there is only 
one term in (\ref{eq_HJ}) contributing to the interaction matrix element. 
Hence the total contribution of all states of the group (iii) to 
$\langle\mbox{Tr}(H_j^2)\rangle$ is $(8/M^3)\,J^2 L(L-1)(M-L)(M-L-1)/4$.

This argumentation does not depend on the choice of initial state giving 
a factor $d$ canceling the factor $1/d$ in (\ref{eq_envar3}). 
Putting this all together, we obtain from (\ref{eq_envar3}) the 
expressions (\ref{eq_DOS_TBRIM}), (\ref{eq_Veff}) and (\ref{eq_a_coeff})
given in the main text for $\sigma$ in terms of the 
effective energy scale $V_{\rm eff}$ and the coefficient $a(M,L)$ which 
measures the global energy rescaling due to finite values of $J/V$. 

\subsection{Numerical verification}

In order to verify numerically the Gaussian density of states with 
the theoretical variance given in (\ref{eq_DOS_TBRIM}) it is more 
convenient to determine the integrated density of states:
\begin{equation}
\label{eq_IDOS}
P(E)=\frac{1}{d}\int_{-\infty}^E \rho(\tilde E)\,d\tilde E\ .
\end{equation}
The prefactor $1/d$ assures the limit $\lim_{E\to\infty} P(E)=1$ 
since $\rho(E)$ is chosen to be normalized to $d$ and not unity. 
In case of an ideal Gaussian density of states, as in (\ref{eq_DOS_TBRIM}), 
we have:
\begin{equation}
\label{eq_IDOS_gauss}
P(E)=
P_{\rm gauss}(E)=\frac{1}{2}\left(1+\mbox{erf}\left(\frac{E}{\sqrt{2}\sigma}
\right)\right)
\end{equation}
with $\mbox{erf}(x)=(2/\sqrt{\pi})\int_0^x \exp(-y^2)\,dy$. 

If $E_m$ represent the numerically computed eigenvalues (of one sample of 
$H_I$ and ordered in increasing order with level number 
$m=0,\ldots,d-1$) the integrated 
density of states is simply obtained by drawing the quantity $z_m=(m+0.5)/d$ 
versus $E_m$ 
which gives the appearance of a rather smooth curve for a sufficiently large 
value of $d$ which can be compared to the expression (\ref{eq_IDOS_gauss}). 
In order to perform a more sophisticated fit analysis we generalize 
(\ref{eq_IDOS_gauss}) to:
\begin{equation}
\label{eq_IDOS_gauss2}
P_k(E)=\frac{1}{2}\left(1+\mbox{erf}\left(q_k(E)/\sqrt{2}\right)\right)
\end{equation}
where $q_k(E)$ is a polynomial of degree $k$. The case $k=1$ with 
$q_1(E)=(E-E_c)/\sigma_{\rm fit}$ corresponds 
to a Gaussian density of states with variance $\sigma_{\rm fit}$ and 
center $E_c$. Choosing larger values of $k>1$ we may analyze deviations with 
respect to the ideal Gaussian distribution. From the practical point of 
view a direct fit of $z_m$ with $P_k(E_m)$ is a bit tricky because it is 
non-linear and it is easier to perform the least-square minimization 
not in the vertical but in the horizontal axis. To do this explicitly 
let, for $0<x<1$, the function $\mbox{inverf}(x)$ be defined as the 
inverse of $\mbox{erf}(x)$ such that $\mbox{erf}(\mbox{inverf}(x))=x$. 
Then we apply the fit $q_k(E_m)=\sqrt{2}\;\mbox{inverf}(2z_m-1)$ which 
is linear in the coefficients of the polynomial $q_k(E)$ and 
provides a unique well defined solution. 

We have applied this fit for the two cases $k=1$ and $k=5$, for 
many different values the ratio $J/V$ covering many orders of magnitude 
and our standard choice $M=16$, $L=7$ with $d=11440$. In all cases 
the hypothesis of an approximate Gaussian density of states is well confirmed 
with a value of $\sigma_{\rm fit}$ confirming the theoretical expression 
in (\ref{eq_DOS_TBRIM}) with an error below 1\%. As an additional 
verification, we have also numerically determined the variance from the trace, 
i.e. the quantity $\sigma_{\rm Tr}^2=(1/d)\mbox{Tr}(H_I^2)=(1/d)\sum_m E_m^2$ 
(the last equality is valid with numerical precision $\sim 10^{-14}$). 
In all cases $\sigma_{\rm Tr}$ also coincides with $\sigma_{\rm fit}$ and the 
theoretical expression with an error below 1\%. 

\begin{figure}[t]
\begin{center}
\includegraphics[width=0.48\textwidth]{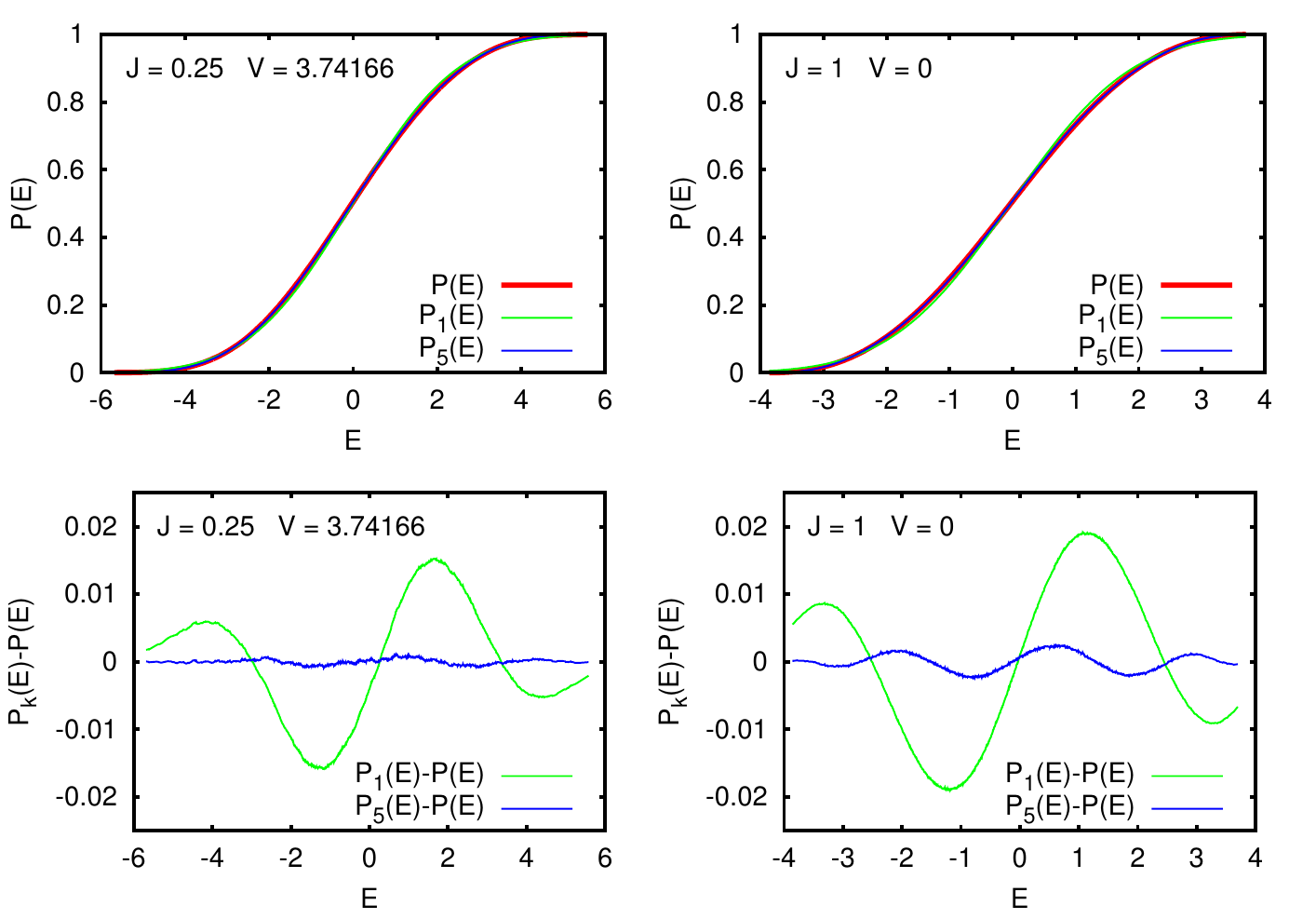}
\end{center}
\caption{\label{fig13}(color online) 
Integrated density of states $P(E)$ of the TBRIM Hamiltonian (\ref{eq_TBRIM}) 
represented by the curve $z_m=(m+0.5)/d$ versus energy level $E_m$ (red curve) 
with $m=0,\,\ldots,\,d-1$ being the level number. The Hilbert space dimension 
is $d=11440$ for $L=7$ particles and $M=16$ orbitals. Shown are the curves 
for one individual spectrum at $V=3.74166$, $J=0.25$ (top left panel) 
and the SYK-case $V=0$, $J=1$ (top right panel). The functions 
$P_k(E)$ correspond to the fit (\ref{eq_IDOS_gauss2}). 
Shown are the cases $k=1$ (green curve) and $k=5$ (blue curve). The case 
$k=1$ corresponds to the (integrated) Gaussian density of states with two 
fit parameters for the width $\sigma_{\rm fit}$ and center $E_c$. 
The fits for $k=1$ provide for $V=3.74166$, $J=0.25$ ($V=0$, $J=1$) 
the values $E_c=-0.008\pm 0.001$ ($-0.032\pm 0.001$) and 
$\sigma_{\rm fit}=1.951\pm 0.001$ ($1.508\pm 0.001$)
giving the ratio $\sigma_{\rm fit}/\sigma=0.9983\pm 0.0006$ 
($0.9952\pm 0.0008$) where $\sigma=1.954$ ($1.516$) is the theoretical value 
obtained from (\ref{eq_DOS_TBRIM}). For comparison the quantity 
$\sigma_{\rm Tr}$, obtained numerically from the trace of $H_I^2$, 
gives for both cases $\sigma_{\rm Tr}=1.947$ ($1.503$). 
In top panels the blue curves for $P_5(E)$ coincide with the red curves 
for $P(E)$ on graphical precision while the green curves for $P_1(E)$ 
are slightly above (below) the red curve for $E>0$ ($E<0$). 
Bottom panels show the difference $P_k(E)-P(E)$ of the fit functions 
with respect to the numerical function $P(E)$ for $k=1$ (green curve) 
and $k=5$ (blue curve) using an increased scale.}
\end{figure}

However, a careful comparison of the numerical curve of $P(E)$ with 
$P_1(E)$ shows small but systematic deviations which can be significantly 
reduced by increasing the degree of the fit polynomial $q_k(E)$. For $k=5$ it 
is already nearly impossible to distinguish the numerical curve from $P_5(E)$ 
on graphical precision. This is clearly illustrated in Fig.~\ref{fig13} where 
we compare the numerical curve $P(E)$ with $P_1(E)$ and $P_5(E)$ 
for the two cases $=3.74166$, $J=0.25$ and $V=0$, $J=1$ (SYK-case). 
In order to see the differences between the two fits it is actually necessary 
to draw the difference of $P(E)-P_k(E)$ in an increased scale 
as it is done in the lower panels of Fig.~\ref{fig13}. 

We attribute the small deviations to the Gaussian 
density of states to the finite values of $M$ and $L$ and also to the fact 
that we used only one numerical sample of $H_J$. Actually, for finite values 
of $J$ it is to our knowledge still an open problem if the average density 
of states of $H_I$ is indeed Gaussian even for the limit $M\to\infty$ and 
$L$ finite (previous analytical results \cite{tezuka,garcia1} 
apply to the SYK-case with Majorana 
fermions that is different from our model at $V=0$ and $J=1$).

\section{Weakly excited initial states}

\label{appb}

The fit procedure using the fit functions (\ref{eq_f11}), (\ref{eq_f01})
to approximate $\rho_{11}(t)$ and $|\rho_{01}(t)|$ are very often quite 
problematic. First 
the non-linear fits with a considerable number of parameters 
depend rather strongly on ``good'' initial values, especially for 
the frequencies $\omega_{1,2}$, for the Levenberg-Marquardt
iteration. Furthermore it is typically necessary to attribute stronger 
weights on the initial times. 
For this we typically perform a first simple exponential 
fit of the survival probability $p(t)=|\langle\psi(0)|\psi(t)\rangle|^2$ 
which provides a smooth simplified decay time which we use to fix 
exponentially decaying weights in time for the more precise fits using the 
fit functions (\ref{eq_f11}), (\ref{eq_f01}). For larger values of the 
couplings strength, typically at $\varepsilon\ge 0.1$, the periodic 
structure with the frequencies $\omega_{1,2}$ also disappears and the fits 
have to be simplified accordingly as mentioned in the main text. 

Even, taking all this into account, for weakly excited initial states, 
with small values of the level number $m$ in (\ref{eq_init_state}), the 
quality of the fits may be rather poor due to the absence of exponential 
decay, presence of a quasi-periodic structure or the effect that after 
an initial decrease $|\rho_{01}(t)|$ re-increases at sufficiently long times. 

\begin{figure}[t]
\begin{center}
\includegraphics[width=0.48\textwidth]{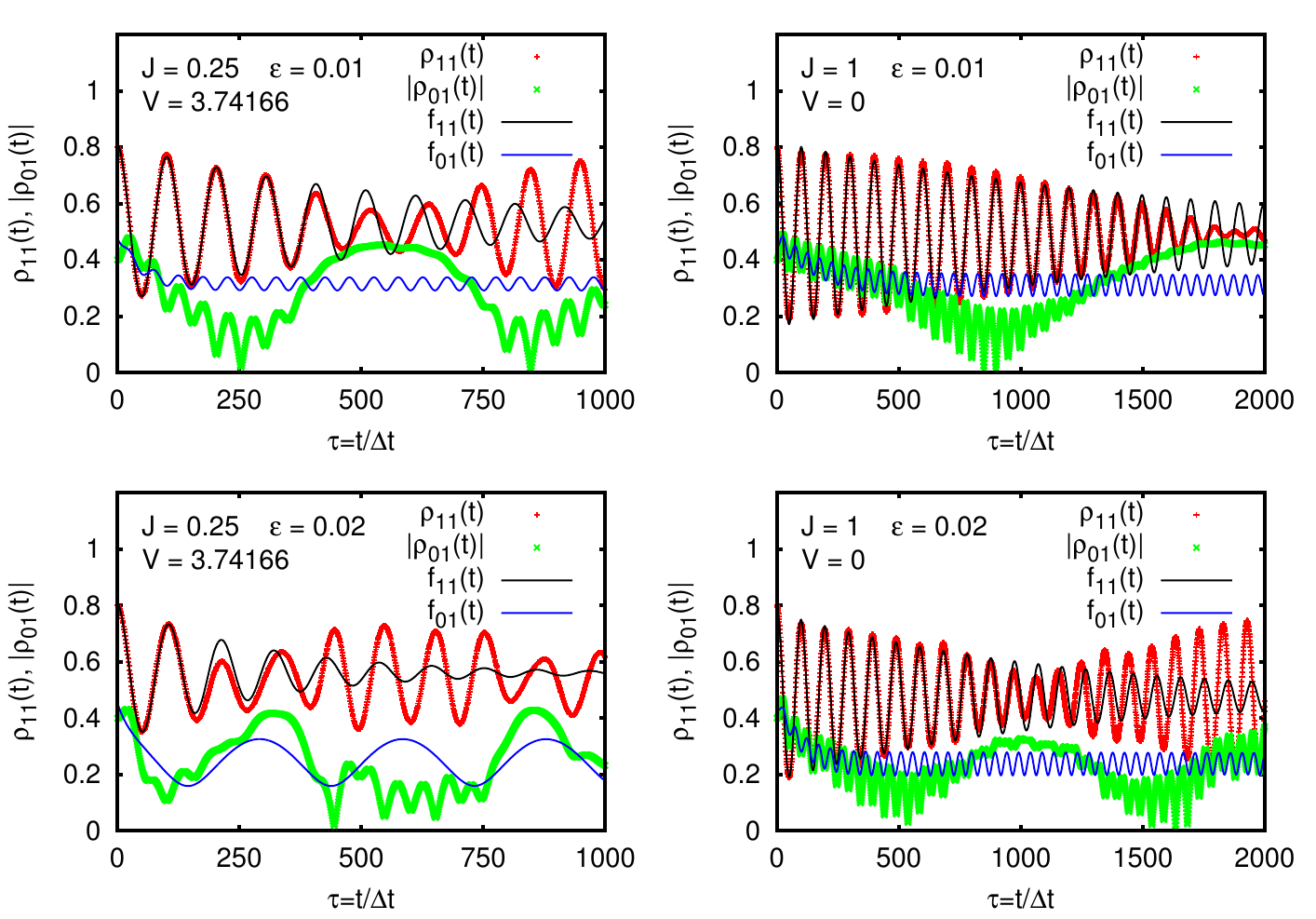}
\end{center}
\caption{\label{fig14}(color online) Time dependence of $\rho_{11}(t)$ 
(red plus symbols), $|\rho_{01}(t)|$ (green crosses) and the two 
fit functions $f_{11}(t)$ (thin black line) and $f_{01}(t)$ (thin blue line), 
defined in (\ref{eq_f11}) and (\ref{eq_f01}), 
for level number $m=7$ of the initial state (\ref{eq_init_state})
and $V=3.74166$, $J=0.25$ ($V=0$, $J=1$) 
in left (right) panels at coupling strength $\varepsilon=0.01$ ($0.02$) 
in top (bottom) panels. 
As in Figs.~\ref{fig2}, \ref{fig3} the time is measured in units of 
$\Delta t$ and the number of particles (orbitals) is $L=7$ ($M=16$). }
\end{figure}

In Fig.~\ref{fig14} we show some examples of this type for the 
level number $m=7$ at our usual standard parameters $V=3.74166$, $J=0.25$ 
or $V=0$, $J=1$ and the coupling strengths $\varepsilon=0.01$ or $0.02$. 
The quantity $\rho_{11}(t)$ exhibits a structure with beats introducing 
a second smaller frequency which is only captured by $f_{11}(t)$ at the 
initial times and even here the deviations due the non-exponential decay 
are quite well visible. For $|\rho_{01}(t)|$ there are fluctuations 
with long correlation times for larger time scales which are not well captured 
by the periodic saturated form of $f_{01}(t)$ at long times. 
In one case at $J=0.25$, $V=3.74166$ and $\varepsilon=0.02$, 
the frequency $\omega_2$ is considerably reduced to fit the 
long range form of $|\rho_{01}(t)|$ but this effect does not reflect 
the physical reality and provides poor values of the two decay rates 
$\Gamma_2$ and $\tilde\Gamma_2$. 

Due to these effects, we do not show any fit functions in Fig.~\ref{fig2}, 
which applies to the level number $m=0$, and in Figs.~\ref{fig4} and 
\ref{fig5} we show the decay rate for the largest level number $m=5720$ 
which is not problematic as can be seen in Fig.~\ref{fig3}. Furthermore 
in Figs.~\ref{fig6} and \ref{fig7}, we only show data points for $m>100$. 

\section{Initial state with negative temperature}

\label{appc}

\begin{figure}[t]
\begin{center}
\includegraphics[width=0.48\textwidth]{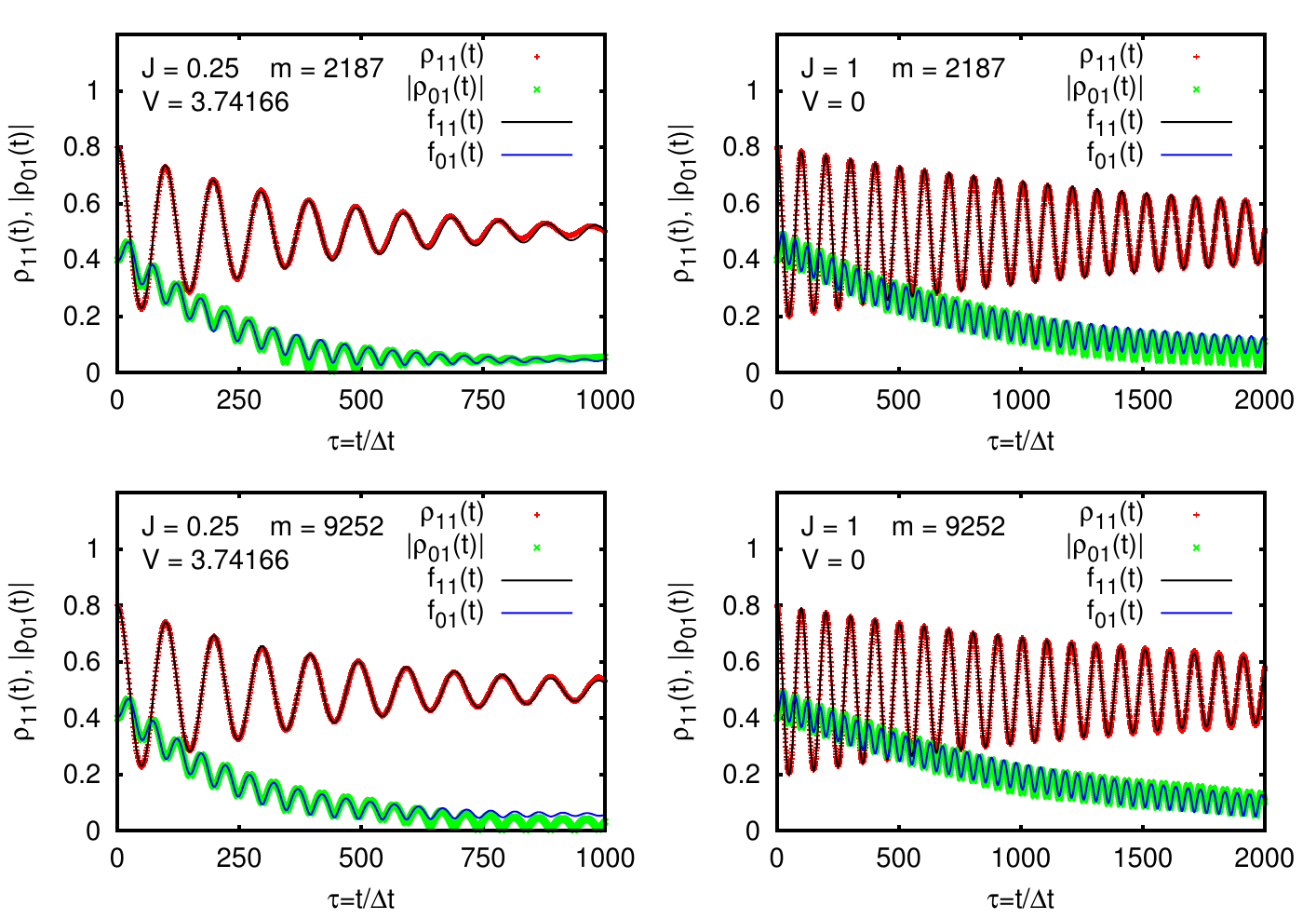}
\end{center}
\caption{\label{fig15}(color online) Time dependence of $\rho_{11}(t)$ 
(red plus symbols), $|\rho_{01}(t)|$ (green crosses) and the two 
fit functions $f_{11}(t)$ (thin black line) and $f_{01}(t)$ (thin blue line), 
defined in (\ref{eq_f11}) and (\ref{eq_f01}), 
for coupling strength $\varepsilon=0.01$ 
and $V=3.74166$, $J=0.25$ ($V=0$, $J=1$) 
in left (right) panels at level numbers $m=2187$ ($9252$) 
in top (bottom) panels. 
The initial state for $V=3.74166$, $J=0.25$ at level number $m=2187$ ($9252$) 
corresponds to the inverse temperature $\beta=0.5443$ ($-0.5424$). 
As in Figs.~\ref{fig2}, \ref{fig3} the time is measured in units of 
$\Delta t$ and the number of particles (orbitals) is $L=7$ ($M=16$). }
\end{figure}

In Fig.~\ref{fig15}
we present the results for qubit relaxation in the regime
of negative temperature (initial state is above the half of energy band width).
Here the dynamical temperature of the in\-itial state
is $T=1/\beta; \beta=-0.5424$. We see that the relaxation
is practically the same as for the initial state
with positive temperature at $\beta=0.5443=1/T$. This effect is due 
to symmetry between negative (positive derivative of the density of states) 
and positive energies (negative derivative of the density of states). 
The former 
correspond to positive and the latter to negative temperatures as can also be 
seen in the bottom panels of Fig.~\ref{fig1}. 

\end{document}